\documentclass[a4paper,11pt]{article}
\pdfoutput=1
\usepackage{jheppub}
\usepackage{xcolor,bm,comment,braket}
\usepackage[T1]{fontenc}
\usepackage{comment}
\usepackage{graphics}

\newcommand{\nn}{\nonumber}
\def\p{\partial}

\newcommand {\beq}{\begin{eqnarray}}
\newcommand {\eeqq}{\end{eqnarray}}

\newcommand {\tr}{{\rm tr}\,}

\newcommand{\SU}{\mathrm{SU}}
\newcommand{\U}{\mathrm{U}}

\def\hsymbu#1{\smash{\lower2.5ex\hbox{\huge$#1$}}}
\makeatletter
\@addtoreset{equation}{section}

\renewcommand{\thefootnote}{\fnsymbol{footnote}}
\makeatother

\begin{document}
\thispagestyle{empty}
\title{String junctions in flag manifold sigma model}

\author[a]{Yuki Amari}
\emailAdd{amari.yuki@keio.jp}
\affiliation[a]{Department of Physics $\&$ 
Research and Education Center for Natural Sciences,
Keio University, 4-1-1 Hiyoshi, Kanagawa 223-8521, Japan}

\author[b,a]{Toshiaki Fujimori}
\emailAdd{toshiaki.fujimori018@gmail.com}
\affiliation[b]{Department of Fundamental Education, Dokkyo Medical University, 880 Kitakobayashi, Mibu, Shimotsuga, Tochigi 321-0293, Japan}

\author[a,c]{Muneto Nitta}
\emailAdd{nitta@phys-h.keio.ac.jp}
\affiliation[c]{
International Institute for Sustainability with Knotted Chiral Meta Matter(SKCM$^2$), Hiroshima University, 1-3-2 Kagamiyama, Higashi-Hiroshima, Hiroshima 739-8511, Japan
}

\author[a]{and Keisuke Ohashi}
\emailAdd{keisike084@gmail.com}

\abstract{
We numerically construct a stable three-string junction of Y-shape in a flag manifold nonlinear sigma model 
on $SU(3)/U(1)^2$.
%containing a four derivative term and a potential. 
}

\maketitle

\setcounter{page}{1}
\setcounter{footnote}{0}
\renewcommand{\thefootnote}{\arabic{footnote}}

%%%%%%%%%%%%%%%%%%%%%%%%%%%%%%%%%%%%%%%%%%%%%%%
%\newpage
\setcounter{tocdepth}{3}
%\tableofcontents
%%%%%%%%%%%%%%%%%%%%%%%%%%%%%%%%%%%%%%%%%%%%%%%%%%%%%%%%%%%
\section{Introduction}

Color confinement in quantum chromodynamics (QCD) is one the most challenging problems in modern physics.
Quarks are confined by color electric flux tubes 
called confining strings. 
Consequently, 
particles observed in nature are hadrons;
mesons are bound states of a quark and an anti-quark confined by a string, 
while baryons consist of three quarks 
confined by three strings possibly joined at a junction. 
Taking a duality, confining strings are mapped to 
${\mathbb Z}_N$ vortices 
in which magnetic fluxes are confined, 
and quarks are mapped to monopoles. 
Thus, monopoles and anti-monopoles are confined by 
color magnetic flux tubes 
as a dual Meissner effect \cite{Nambu:1974zg,Mandelstam:1974pi}.
When one pulls a constituent quark in a meson, a confining string is elongated and one would observe a pair creation of quark and anti-quark and the string breaking, resulting in two mesons.
If one does so for a baryon, 
three-string junction will be seen
until the string breaking occurs.
A lattice QCD simulation of a bound state of three heavy quarks clearly shows a Y-junction of 
three confining strings 
\cite{ 
Takahashi:2000te,
Takahashi:2002bw,
Bali:2000gf}. 
More generally, $N$ strings join at a junction 
called a baryon vertex 
in an $\SU(N)$ Yang-Mills theory. 
Recently, quantum fluctuations around the string junction have been discussed and a constraint on the string junction mass has been proposed \cite{Komargodski:2024swh}.
In string theory, confining strings are  identified with fundamental strings 
\cite{Donagi:1995cf,Polchinski:2000uf}.
Recently, 
confining strings are also proposed as 
cosmic strings \cite{Yamada:2022imq}.

On the other hand, 
Faddeev and Niemi proposed that glueballs can be represented by knotted 
strings \cite{Faddeev:1996zj}.
To this end, they proposed by using 
the so-called Cho-Faddeev-Niemi or Cho–Duan–Ge–Faddeev–Niemi–Shabanov decomposition \cite{Cho:1979nv,Cho:1980vs,Duan:1979ucg,Faddeev:1998eq,Faddeev:1998yz,Faddeev:1999cj,Shabanov:1999uv} (for a review, see Ref.\,\cite{Kondo:2014sta}) that 
$\SU(2)$ Yang-Mills theory is reduced to
the Faddeev-Skyrme model, 
an $O(3)$ nonlinear sigma model with 
a four derivative term \cite{Faddeev:1975tz,Faddeev:1976pg}.\footnote{
However, there is also an objection 
to this claim \cite{Evslin:2010sb}.
}
%\cite{Faddeev:1996zj,Faddeev:1998eq,Faddeev:1998yz,Faddeev:1999cj}, see also \cite{Kondo:2014sta}. 
The target space is $S^2$ and 
topological lumps supported by the second homotopy group 
$\pi_2 (S^2) \simeq {\mathbb Z}$ are 
identified with confining strings. 
A straight string was discussed in 
Refs.\,\cite{Ferreira:2011mz}. 
Furthermore, this model admits Hopfions topologically characterized by $\pi_3 (S^2) \simeq {\mathbb Z}$, 
which are closed lump strings 
\cite{Battye:1998pe,Battye:1998zn,Hietarinta:1998kt,PhysRevD.62.081701,Kobayashi:2013xoa}, 
see Refs.\,\cite{Radu:2008pp,Shnir:2018yzp} for a review.
Faddeev and Niemi proposed 
that these Hopfions can be identified with glueballs.

More realistic case for QCD is 
the gauge group $G=\SU(3)$.
For the gauge group $G=\SU(N)$, the decomposition results in 
a flag manifold \cite{Cho:1980vs,Faddeev:1998yz,Kondo:2014sta}
\begin{align}
F_{N-1} \simeq  \SU(N)/{\U(1)^{N-1}}.
\end{align}
The flag manifold sigma models have been recently studied in various contexts in high energy physics and 
condensed matter physics 
\cite{Bykov:2011ai,Bykov:2012am,Bykov:2014efa,Bykov:2015pka,Bykov:2019jbz,Bykov:2019vkf,Hongo:2018rpy,Tanizaki:2018xto,Ohmori:2018qza,Lajko:2017wif,Wamer:2019oge,smerald2013theory,
PhysRevA.93.021606,Amari:2017qnb,Amari:2018gbq,Wamer:2020inf,Kobayashi:2021qfj,Takahashi:2021lvg}, 
see Ref.\,\cite{Affleck:2021ypq} for a review:
spin chains \cite{Bykov:2011ai,Bykov:2012am,Lajko:2017wif,Wamer:2019oge,Wamer:2020inf},
flag manifold sigma model on ${\mathbb R}\times S^1$ \cite{Hongo:2018rpy},
anomaly and topological $\theta$ term \cite{Tanizaki:2018xto,Kobayashi:2021qfj},
world-sheet theories of composite non-Abelian vortices \cite{Eto:2010aj,Ireson:2019gtn}, 
and a non-Abelian vortex lattice \cite{Kobayashi:2013axa}.
The flag manifold sigma models admit 
several types of topological lumps because of the second homotopy group
\begin{align}
\pi_2(F_{N-1}) \simeq \pi_1 [\U(1)^{N-1}] \simeq {\mathbb Z}^{N-1}. 
\label{eq:lumps}
\end{align}
Various properties of the topological 
lumps have been elucidated in Refs.\,\cite{negreiros1988,Bykov:2015pka,PhysRevA.93.021606,Amari:2017qnb,Fujimori:2023wkd}.  
In our previous paper 
\cite{Fujimori:2023wkd}, we exhausted 
 Bogomol'nyi-Prasad-Sommerfield (BPS) lumps in supersymmetric K\"ahler flag sigma models 
\cite{Bando:1983ab,Bando:1984cc,Bando:1984fn,Itoh:1985ha,Itoh:1985jz,Nitta:2003dv} 
 and determined their moduli space 
 in terms of the moduli matrix \cite{Isozumi:2004vg,Eto:2004rz,Eto:2005yh,Eto:2006cx,Eto:2006pg}.
When we regard the flag manifold sigma model 
as a low-energy theory of the $\SU(N)$ Yang-Mills theory 
along the line of Faddeev and Niemi \cite{Kondo:2014sta}, 
Hopfions in the $F_2$ Faddeev-Skyrme model
were discussed in Ref.\,\cite{Amari:2018gbq}.
In this case, to justify 
the $F_2$ Faddeev-Skyrme model as 
a certain low-energy theory of the $\SU(3)$ Yang-Mills theory,
the model should admit a three-string junction, 
which is the main target of this paper.

In this paper, we show that a stable string junction is indeed present in the flag manifold sigma model with a four derivative term and a potential term. 
By using gradient descent method, we numerically construct a three-string junction of 
a Y-shape in the $F_2 \simeq \SU(3)/{\U(1)^{2}}$ sigma model, which could be relevant for the $\SU(3)$ 
Yang-Mills theory.
%More specifically, in this paper, for the purpose to simulate a three-string junction elongated to infinity in ${\mathbb R}^3$, we compactify the space to a torus $T^3$ with twisted boundary conditions\footnote{The twist on the boundary conditions is not essential but just introduced due to the technical requirement that the BPS solution be consistent with those conditions.}, so that one can perform numerical simulations in a finite box, and sheets of crystals composed of the string junctions are numerically constructed. % by using the relaxation method.
Our interest is in a single junction in $\mathbb{R}^3$ space. However, realizing an isolated single junction in a finite space is complicated and difficult. Therefore, we place a junction/anti-junction pair in a torus, impose periodic boundary conditions to prevent their annihilation, and eliminate interaction effects by analyzing the torus-size dependence of the configuration energy.
The technical key point of the
gradient descent method is choosing a topologically correct initial state. 
As a simple and reliable choice for that,  we adopt a configuration deformed from a BPS solution in a K\"ahler $F_2$ nonlinear sigma model.

This paper is organized as follows.
In Sec.\,\ref{sec:model} 
we define the $F_{N-1}$ Faddeev-Skyrme model.
In Sec.\,\ref{sec:BPS-lump}, we present BPS lump solutions 
in the $F_2$ sigma model based on Ref.\,\cite{Fujimori:2023wkd}. 
In Sec.\,\ref{sec:junction}, we numerically construct 
a stable string junction of Y-shape in 
the $F_2$ 
%\simeq  SU(3)/{U(1)^{2}}$ 
sigma model 
on a three dimensional torus $T^3$ and we also discuss the instability of the strings caused  by junction pair production. 
Sec.\,\ref{sec:summary} is devoted to a summary and discussion.
In Appendix \ref{sec:parametrization},
we give a relation between 
the parametrizations of the model used in this paper and the original one by Faddeev and Niemi.
In Appendix \ref{sec:generalFN-1}, we give comments on the general $F_N$ sigma model.
In Appendix \ref{appendix:NumericalCalculation},  
we give some details of our numerical calculations.
In Appendix \ref{sec:otherBC}, we examine cases with other boundary conditions.
%%%%%%%%%%%%%%%%%%%%%%%%%%%%%%%%%%%%%%%%%%%%%%%%%%%%%%%%%%%%%%%%%%%%%%%%%%%%%
%%%%%%%%%%%%%%%%%%%%%%%%%%%%%%%%%%%%%%%%%%%%%%%%%%%%%%%%%%%%%%%%%%%%%%%%%%%%%%%%
%%%%%%%%%%%%%%%%%%%%%%%%%%%%%%%%%%%%%%%%%%%%%%%%%%%%%%%%%%%%%%%%%%%%%%%%%%%%%%%%%%%%%%%

%%%%%%%%%%%%%%%%%%%%%%%%%%%%%%%%%%%%%%%%%%%%%%%%%%%%%%%%%%%%%%%%%%%%%%%%%%%%
\section{\texorpdfstring{$F_{N-1}$}{F(N-1)} Faddeev-Skyrme model} \label{sec:model}

In this section, we present the model and provide an overview of the topological lumps in the model.

\subsection{The model}
In this paper, we use
the Minkowski metric convention $\eta_{\mu\nu}={\rm diag} (-1,1,1,1)$. 
The model we consider in this paper is a $3+1$ dimensional theory called the $F_{N-1}$ Faddeev-Skyrme model. For the convenience of numerical calculations, we take a three-dimensional torus $T^3$ as the base space. The Lagrangian is defined by
\begin{align}
-{\cal L} \ = \ \sum_{i=1}^{N} \left( \frac{f^2}{4} \tr[\p_\mu P_i \p^\mu P_i] + \frac{1}{8g^2} 
%\sum_{i=1}^{N} 
F^i_{\mu\nu}F^{i,\mu\nu} \right) +V,
\label{eq:Lagrangian_P}
\end{align}
where $\{P_i=P_i(x^\mu)| i=1,2,\dots,N\}$
 is a set of projection matrices of order $N$ satisfying
\begin{align}
P_i=(P_i)^\dagger, \quad  P_iP_j=\delta_{ij} P_i,  \quad \sum_{i=1}^{N}P_i ={\bf 1}_{N},\quad \tr[P_i]=1,
\end{align}
$F_{\mu\nu}^j$'s are defined by 
\begin{align}
%A_\mu^j \equiv i\tr[p_j \p_\mu U U^\dagger ], \hspace{10mm}
F_{\mu\nu}^j \equiv 
%\p_\mu A_\nu^j -\p_\nu A_\mu^j
-i \tr\left[P_j [\p_\mu P_j,\, \p_\nu P_j]\right],
\end{align}
and $V$ is a potential term given below.
In Eq.~\eqref{eq:Lagrangian_P},
the first term is the $F_{N-1}$ sigma model, the second term with four derivatives is called the Skyrme term, and the last term is a potential.
The %four-derivative 
Skyrme term %(the second term) 
and the potential $V$ are introduced to avoid a subtle problem on the stability of string junctions (see Sec.\,\ref{sec:junction}).     
%necessary to stabilize the thickness of the strings discussed below. 
The Skyrme term was present in the proposal of Faddeev and Niemi 
\cite{Faddeev:1998yz}. 
See Appendix~\ref{sec:parametrization} for the equivalence between the model in 
Eq.\,(\ref{eq:Lagrangian_P})
and one used by  
Faddeev and Niemi 
\cite{Faddeev:1998yz} 
in which a potential is not considered.

%The target manifold $F_{N-1}$ is described by a set of $N$ projection matrices $\{P_i| i=1,2,\dots,N\}$ satisfying
%\begin{align}
%P_i=(P_i)^\dagger, \quad  P_iP_j=\delta_{ij} P_i,  \quad \sum_{i=1}^{N}P_i ={\bf 1}_{N},\quad \tr[P_i]=1.
%\end{align}
The set of $\{P_i\}$ can always be expressed with a unitary matrix $U=U(x^\mu) \in \U(N)$ as
\begin{align}
P_i=U^\dagger p_i \, U \quad{\rm for~} i=1,2,\cdots,N, \label{eq:PtoU}
\end{align}
where $p_i$'s are the reference projection matrices  defined by
\begin{align}
(p_i)^a_b=\delta_{ia}\delta_{ib}. 
\end{align}
In this expression, 
there exists a redundancy of $\U(1)^{N}$ acting on $U$ as
 \begin{align}
U \sim U'= e^{i \Theta } U \quad  {\rm  with~~~} \Theta= \sum_{i=1}^{N} \theta_i p_i \quad \theta_i=\theta_i(x^\mu) \in \mathbb R/2\pi\mathbb Z \simeq S^1.
\end{align}
This $\U(1)^{N}$  redundancy is a hidden local symmmetry and 
$F_{\mu\nu}^i$'s are nothing but 
field strengths of 
(composite) $\U(1)^{N}$ gauge field $A_\mu^i$: 
\begin{align}
A_\mu^j \equiv i\tr[p_j \p_\mu U U^\dagger ], \hspace{10mm}
F_{\mu\nu}^j = \p_\mu A_\nu^j -\p_\nu A_\mu^j.\label{eq:gaugefield}
%=-i \tr\left[P_j [\p_\mu P_j,\, \p_\nu P_j]\right], 
\end{align}

In the following, we set the potential term as 
\begin{align}
V = \frac{\mu^2}{4} \sum_{i=1}^N{\rm tr} \big[ ( p_i - P_i )^2 \big].
\end{align}
This potential explicitly breaks the $\U(N)$-flavor symmetry to $\U(1)^{N}$ and hence 
the fluctuations around the vacuum $U={\bf 1}$ have the mass $\mu/f$. 
In addition to the unbroken $\U(1)^{N}$ symmetry, this model has an $S_N$ symmetry which permutes the projection matrices $P_i \leftrightarrow P_j$.\footnote{An element $\sigma$ of $S_N$ acts on the unitary matrix $U$ as ${\cal P}_\sigma^{-1}U{\cal P}_\sigma$ with the permutation matrix ${\cal P}_\sigma$. }
This $S_N$ symmetry is particular for the model in Eq.\,\eqref{eq:Lagrangian_P}  
and does not exist in the $F_{N-1}$ flag sigma model with more general coefficients, 
given in Appendix \ref{sec:generalFN-1}. 

\subsection{Topological lump strings}

Here, we discuss topological lumps 
corresponding to the second homotopy group $\pi_2(F_{N-1}) \simeq \mathbb Z^{N-1}$ in Eq.\,(\ref{eq:lumps}).
According to the topological charges, 
the topological sectors can be classified by the following topological invariant
\begin{align}
 {\mathbf m}=(m_1,m_2,\cdots,m_{N}):\quad  m_j  \equiv \frac1{4\pi i} \int_\Sigma dx^\mu\wedge dx^\nu  F_{\mu\nu}^j  \quad \in\mathbb Z, \label{eq:topcharge}
\end{align}
where $\Sigma$ is a two-dimensional plane embedded into the base space $T^3$.
Note that this topological invariant has $N-1$ degrees of freedom since it must satisfy the constraint\footnote{
It is obvious from Eq.\eqref{eq:gaugefield}   that the overall $U(1)$ gauge field is a pure gauge as
\begin{align*}
\sum_{i=1}^N A_\mu^i=i \p_\mu \log \det U. 
\end{align*}
}
\begin{align}
\sum_{i=1}^{N} m_i=0 
\hspace{10mm} 
\left( \because \sum_{i=1}^{N} F^i_{\mu\nu}=0 \right).
\end{align}
This topological invariant guarantees the existence of string-like topological solitons, which are orthogonal to the plane $\Sigma$.  
Among such string-like objects, 
an elementary one carries a charge 
\begin{align}
 {\mathbf m}= {\mathbf m}^{\langle i,j \rangle } \equiv (0,\dots, 0, \stackrel{i}{+1},0, \cdots,0,\stackrel{j}{-1},0,\dots ),
\end{align}
We call the string-like object with this topological charge $\langle i,j \rangle $-string. 
Each string has a direction: for example, a $\langle j,i \rangle $-string is a $\langle i,j \rangle $-string extending in the opposite direction.
Thus, a pair of parallel $\langle i,j \rangle $- and  
$\langle j,i \rangle $-strings can be annihilated:
\begin{align}
{\mathbf m}^{\langle i,j \rangle }+{\mathbf m}^{\langle j,i \rangle }=0.
\end{align}
The composite state of an $\langle i,j \rangle $- and $\langle j,k \rangle $-strings orthogonal to the plane $\Sigma$ 
has the same topological charge with 
an $\langle i,k \rangle $-string:
%that penetrates in the same way:
\begin{align}
{\mathbf m}^{\langle i,j \rangle }+{\mathbf m}^{\langle j,k \rangle }={\mathbf m}^{\langle i,k \rangle }.
\end{align}
All elementary strings have the same tensions, $T_{\langle i,j\rangle }=T$, thanks to the $S_N$ symmetry in our model.
Therefore, a single ${\langle i,k \rangle}$-string
is energetically more stable than 
two separated ${\langle i,j \rangle }$- and ${\langle j,k \rangle }$-strings:  $T_{\langle i,k\rangle }=T
<2T=T_{\langle i,j\rangle}+T_{\langle j,k\rangle }$, which means there exists a binding energy between the ${\langle i,j \rangle }$- and ${\langle j,k \rangle }$-strings. 
%\footnote{ 
%There is no need to discuss the details of the interaction between the two strings here. 
%We just assume that single elementary string is globally stable. That is,
%a energy of the composite state, $V(l)$ with a distance $l$ is assumed only to have the following properties: 
%$T<V(l)<\infty$ for $l>0$, $V(0)=T$ and $V(l)\approx 2T$ for large $l$.
%When considering a string junction, 
%the contribution from three strings well away from each other dominates in the total energy, 
%so no details of the potential $V(l)$ are relevant to whether or not a string junction is formed.
%If a large potential barrier exists, it should appear as a positively large "junction mass", which will be defined in %Eq.\,\eqref{eq:junctionmass}.
%} 

\begin{figure}[ht]
\centering
\includegraphics[width=6cm]{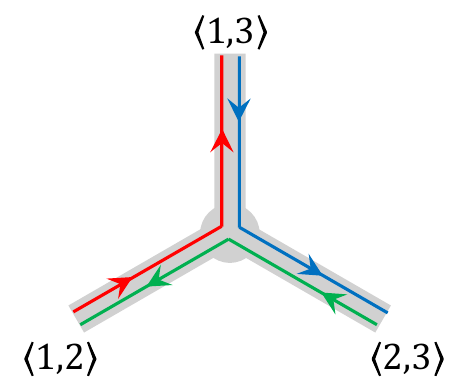}
\caption{ \label{fig:JunctionImage} %Image for 
Schematic picture of a string junction in the $N=3$ case. %$F_2$ sigma model.
}
\end{figure}

Based on these facts, 
it is quite natural to expect that 
three ${\langle i,j \rangle }$-, ${\langle j,k \rangle }$- and  ${\langle k,i \rangle } $-strings 
extending from three directions meet at a single point and form a string junction as illustrated in Fig.\,\ref{fig:JunctionImage}.
Due to the balance of forces, all three angles between 
the strings in this junction should be $2\pi/3$. 
The aim of this paper is to show that this configuration indeed exists as a stable solution of the equation of motion.
%%%
%%%

\section{Initial configurations}\label{sec:BPS-lump}

In this section, we prepare initial configurations 
for iterative numerical simulations 
to construct a string junction solution 
discussed in the next section. 
It is sufficient to prepare initial configurations 
of the same topology with expected string junctions, 
since 
the topology of the configuration remains unchanged
in the numerical process. 
%Therefore, the initial configuration should have the %correct topology to lead to the string junction. 
Although any configuration 
%is applicable,To obtain an initial configuration 
which is topologically equivalent to the string junction is fine, 
here we consider a deformed model 
which can be obtained from our model by continuous change of parameters: 
 a K\"ahler flag manifold sigma model \cite{Bando:1983ab,Bando:1984cc,Bando:1984fn,Itoh:1985ha,Itoh:1985jz,Nitta:2003dv} 
 without the Skyrme term and the potential term ($1/g^2=\mu^2=0$). 
In such a model, analytic solutions 
of BPS strings are available. 
The initial configurations that we provide 
are solutions to the equation of motion of the deformed model, giving good ansatz 
among all configurations with the same topology.
%we turn off the Skyrme term and potential term in this section: $1/g^2=\mu^2=0$. 
%We also deform the target space of the theory to a K\"ahler flag manifold  \cite{Bando:1983ab,Bando:1984cc,Bando:1984fn,Itoh:1985ha,Itoh:1985jz,Nitta:2003dv} to allow for a BPS string solution in order to obtain a topologically equivalent initial configuration, and consider a slice of the base space with constant $x_2$, where a string object becomes just a lump in 2+1 dimensions. 
In this section, we provide an overview the construction of a general BPS solution composed of a $\langle 1,2\rangle $-lump and a $\langle 2,3\rangle$-lump in the K\"ahler flag manifold sigma model.

%%%%%%%%%%%%%%%%%%%%%%%%%%%%%%%%%%%%%%%%%%%%%%%%%%%%%%%%%%%%%%%%%%%%
\subsection{Non-K\"ahler and K\"ahler \texorpdfstring{$F_2$}{F(2)} sigma models}
Now let us focus on the case of  the $F_2$ sigma model ($N=3$). Here, we consider the general $F_2$ sigma model, which has three parameters.\footnote{
The general $F_{N-1}$ sigma model has $N(N-1)/2$ parameters as shown in Appendix \ref{sec:generalFN-1}.}
%In an $N$-dimensional subspace of the parameter space, 
The model takes the form of three copies of the $\mathbb CP^2$ sigma model with some constraints
\begin{align}
-{\cal L}_{\sigma\text{-model}}%K
=
\sum_{i=1}^{3} \frac{r_i}2 \tr[\p_\mu P_i \p^\mu P_i]=\sum_{i=1}^{3} r_i K_{\rm FS}(w_i, w_i^\dagger )\quad {\rm with} \quad P_i = \frac{w_i^\dagger w_i}{|w_i|^2} . 
\label{eq:CPLsum}
\end{align}
Here $w_i\in \mathbb C^{3}\backslash\{{\bf }0\}$ $(i=1,2,3)$ are row vectors representing the homogeneous coordinates with the equivalence relation $w_i \sim \lambda_i w_i~(\lambda_i \in \mathbb C\backslash \{0\})$.  
They are not independent and must satisfy the orthogonality constraints 
\begin{align}
w_i \cdot w_j^\dagger = 0 \quad \mbox{for} \quad j \not = i.
\end{align}
$K_{\rm FS}$ is the kinetic term 
with the Fubini-Study metric
\begin{align}
K_{\rm FS}(w, w^\dagger )\equiv \frac1{|w|^2}
\partial^\mu w \left({\bf 1}-\frac{ w^\dagger w}{|w|^2}\right) \partial_\mu w^\dagger, \hspace{10mm}
w \in \mathbb C^{3} \backslash \{ 0 \}.
\end{align}
The coefficients $r_i~(i=1,2,3)$ must satisfy inequalities
\begin{align}
     r_i+r_j >0  \quad{\rm for~} j\not=i.
\end{align}
Note that they can take negative values as long as these inequalities are satisfied. 
In terms of $w_i$, the topological number $m_i$, 
defined in Eq.\eqref{eq:topcharge}, 
takes the following form seen in the $\mathbb CP^{2}$ sigma model: 
\begin{align}
m_i =\frac1{2\pi i} \int_{\Sigma} d 
\left( \frac{d w_i\cdot w_i^\dagger}{|w_i|^2} \right). 
%=\frac{1}{2\pi i}\oint_{\p \Sigma_{\rm in}} \frac{d \lambda_i}{\lambda_i}
\label{eq:mbylambda}
\end{align}

%In this case, Eq.\,\eqref{eq:CPLsum} is the most general case %kinetic term since
%because $N(N-1)/2=3$ for $N=3$. 
Note that $F_2$ is a complex manifold with
three inequivalent complex structures. 
For each choice of the complex structure, 
$F_2$ becomes a K\"ahler manifold 
by setting one of $r_i$ to be zero.
To express BPS solutions, 
it is convenient to use the complex coordinates $\{\phi_1,\phi_2,\phi_3\}$\footnote{The complex coordinates $\{\phi_1,\phi_2,\phi_3\}$ are the parameters contained in the coset matrix as
\begin{align}
U = \hat h^{-1} e^{\Phi}, \quad \Phi =
\left(\begin{array}{ccc}
0 & \phi_3 & \phi_2 \\
0 & 0 & \phi_1 \\
0 & 0 & 0
\end{array}\right), \notag
\end{align}
where $U$ is the unitary matrix appearing in Eq.\,\eqref{eq:PtoU} and $\hat h$ is the lower triangular matrix which can be determined from $\hat h \hat h^\dagger =e^{\Phi}e^{\Phi^\dagger}$ up to $\U(1)^3$.}
such that $w_i$ are given as
\begin{align}
w_1 \sim (1\,,\phi_3\,,\phi_2^+),  \hspace{5mm} 
w_2 \sim (-(\phi_3')^*, 1\,, \phi_1'), \hspace{5mm}
w_3\sim (-(\phi_2^-)^*,-\phi_1^*\,,1\,)
\end{align}
where $\phi_i^*$ stands for the complex conjugate of $\phi_i$ and
\begin{align}
\phi_2^\pm \equiv \phi_2\pm \frac12 \phi_3\phi_1,\quad  \phi_1'\equiv \frac{\phi_1-\phi_3^* \phi_2^-}{1+(\phi_2^+)^* \phi_2^-},\quad \phi_3'\equiv\frac{\phi_3+ \phi_1^* \phi_2^+}{1+\phi_2^+( \phi_2^-)^*}.
\end{align}
We can confirm that if one of the parameters $r_i$ 
is set to zero, the target manifold becomes a K\"ahler manifold, 
otherwise it is not a K\"ahler manifold. %, and moreover, it is not even Hermitian.
Note that if two of $r_i$ are set to zero, 
the target space reduces to $\mathbb{C}P^2$.

%%%%%%%%%%%%%%%%%%%%%%%%%%%%%%%%%%%%%%%%%%%%%%%%%%%%%%%%%%%%%%%%%%%%%%
\subsection{BPS lumps in the \texorpdfstring{$F_2$}{F(2)} sigma model on \texorpdfstring{$T^2$}{T(2)}}\label{sec:BPS}
\def\sn{{\rm sn}}
\def\sc{{\rm sc}}
\def\sd{{\rm sd}} 
%Here we temporally set $1/g^2=\mu^2=0$ and 
Here, we consider strings parallel to the $x_2$-axis, 
which can be viewed as lump solutions localized on the perpendicular 2D plane $\Sigma \simeq T^2\subset T^3$. 
Therefore, in this subsection, we consider the $2+1$-dimensional theory and regard $\Sigma \simeq T^2$ as the base space. 
We set $r_2=0$ to obtain a K\"ahler $F_2$ sigma model admitting BPS lump solutions. 
In this case, there is no interaction between $\langle 1,2 \rangle$-lumps and $\langle 2,3 \rangle$-lumps.
BPS lump solutions are given by holomorphic maps from $\mathbb C$ to $F_2$ which are represented by meromorphic functions $(\phi_1(z),\phi_2(z),\phi_3(z))$ of $z=x_1+ix_3 \in \mathbb C$.

Let us construct a single BPS lump solution on $T^2$ by embedding a single $\mathbb C P^1$ lump solution into one of the complex coordinates $\{\phi_1,\phi_2,\phi_3\}$ of $F_2$. 
Any $\mathbb C P^1$ BPS lump solution is given by a meromorphic function and a single lump solution has a single pole (and a single zero).
However, all doubly periodic meromorphic functions 
must have at least two poles 
(and two zeros) in their fundamental domains. 
To obtain a single BPS lump solution, 
let us define $T^2$ as
$T^2=\mathbb C/\sim$ with $z \sim z+p L_1+ iqL_3~(p,q \in \mathbb Z)$ by dividing the fundamental domain into two domains and allowing twisted periodicity on 
$\{\phi_1,\phi_2,\phi_3\}$ as
\begin{align}
(\,\phi_1(z+L_1)\,,\,\phi_2^\pm(z+L_1)\,,\,\phi_3(z+L_1)\,)=(+\phi_1(z),-\phi_2^\pm(z),-\phi_3(z)), \label{eq:phiperiodicity1} \\
(\phi_1(z+iL_3),\phi_2^\pm(z+iL_3),\phi_3(z+iL_3))=(-\phi_1(z),-\phi_2^\pm(z),+\phi_3(z)).
\label{eq:phiperiodicity2}
\end{align}
To construct solutions, 
it is convenient to use the Jacobi's elliptic functions $\sn(u)=\sn(u;k),\, \sc(u)=\sc(u;k)$ and $\sd(u)=\sd(u;k)$, 
which have different twisted periodicity given by
\begin{align}
%(\,\sc(u+2\,K\,;k)\,,\,\sd(u+2\,K\,;k)\,,\,\sn(u+2\,K\,;k)\,) = (+\sc(u;k),-\sd(u;k),-\sn(u;k)), \\
%(\sc(u+2iK';k),\sd(u+2iK';k),\sn(u+2iK';k)) = (-\sc(u;k),-\sd(u;k),+\sn(u;k)), 
(\,\sc(u+2\,K\,)\,,\,\sd(u+2\,K\,)\,,\,\sn(u+2\,K\,)\,) = (+\sc(u),-\sd(u),-\sn(u)), \\
(\sc(u+2iK'),\sd(u+2iK'),\sn(u+2iK')) = (-\sc(u),-\sd(u),+\sn(u)), 
\end{align}
where $K=K(k) (K'=K'(k))$ is the complete elliptic integral of the first kind
\begin{align}
K(k) = \int_0^{\frac{\pi}{2}} \frac{d\theta}{\sqrt{1-k^2 \sin^2\theta}}, \hspace{10mm} K'(k) = K(\sqrt{1-k^2}).
\end{align}
Setting the elliptic modulus $k$ so that $K(k)/K'(k) = L_1/L_3$, we can find $L_*\in \mathbb R_{>0}$ such that $K=L_1/(2L_*)$ and $K'=L_3/(2L_*)$. 
Then, by identifying the coordinates as $u=z/L_*$, 
we can show that each of $(\sn(u;k)^{-1},\sc(u;k)^{-1},\sd(u;k)^{-1})$ is a function that satisfies the twisted boundary conditions \eqref{eq:phiperiodicity1}-\eqref{eq:phiperiodicity2} and has a single pole at $z=0$ on $T^2$. 
Therefore, using these functions, we can write down single BPS lump solutions. 

The general solution for a single $\langle 1,2 \rangle$-lump is given by 
\begin{align}
\phi_3=\frac{c_3}{{\rm sn}(u-u_3)}, \quad \phi_2^\pm=\phi_1=0 \quad {\rm with} \quad {\bf m}={\bf m}^{\langle 1,2 \rangle} \equiv (1,-1,0),
\label{eq:singlelump3}
\end{align}
where $\phi_3$ has only one pole at $u=u_3$.
Similarly, the general solution for a single ${\langle 2,3 \rangle }$-lump is given by 
\begin{align}
\phi_1=\frac{c_1}{{\rm sc}(u-u_1)}, \quad \phi_2^\pm=\phi_3=0 \quad {\rm with} \quad {\bf m}={\bf m}^{\langle 2,3 \rangle} \equiv (0,1,-1).\label{eq:singlelump1}
\end{align}
Here, $c_1$ and $c_3$ are dimensionless moduli parameters. 
The quantity $|c_i|L_*$ roughly gives the size of each lump.  
We assume that $|c_i|$ is sufficiently smaller than 1 so that the energy density profile of each lump is localized around $u=u_i$ (the poles of $\phi_i$).\footnote{ 
Conversely, if $|c_i|$ is sufficiently larger than 1, we observe an object of size $|c_i|^{-1}L_*$ localized around
zero of $\phi_i$.} 

Next, let us consider composite states of 
${\langle 1,2 \rangle }$ and ${\langle 2,3 \rangle }$-lumps 
carrying charge ${\bf m}={\bf m}^{\langle 1,2 \rangle }+
{\bf m}^{\langle 2,3 \rangle }$.
Eq.,\eqref{eq:mbylambda} implies that in order to have topological charges ${\bf m}={\bf m}^{\langle 1,2 \rangle }+
{\bf m}^{\langle 2,3 \rangle }$, $(\phi_3,\phi_2^+)$ and
$(\phi_1,\phi_2^-)$ should each have only one pole at $u=u_3$ and $u=u_1$, respectively.
Therefore, the general solution is given by\footnote{
Here the coefficients in the above $\phi_2^\pm$ are uniquely determined as follows. 
The poles of $\phi_2^+$ and $\phi_2^-$ are located at the different points, 
although $\phi_2^\pm$ are not independent of each other
and must satisfy a relation
\begin{align}
0=f(u)\equiv \phi_2^-(u)-\phi_2^+(u)+\phi_3(u)\phi_1(u).\nonumber
\end{align}
Therefore, the coefficients in $\phi_2^\pm$ must be determined so that the two poles in $f(u)$ cancel out. 
Once the coefficients are determined in this way, $f(u)$ becomes a constant function due to the property of elliptic functions, and furthermore, the twisted periodicity of $f(u)$ automatically requires that $f(u)=0$.  
}
\begin{align}
\phi_3=\frac{c_3}{{\rm sn}(u-u_3)},\qquad&\phi_2^+ = \frac{c_1c_3}{{\rm sc}(u_3-u_1)}\frac{1}{{\rm sd}(u-u_3)},\qquad \nn\\
\phi_1=\frac{c_1}{{\rm sc}(u-u_1)}, \qquad &\phi_2^- = \frac{c_1c_3 }{{\rm sn}(u_3-u_1)}\frac{1}{{\rm sd}(u-u_1)}.\qquad \label{eq:compositestate}
\end{align}

By setting  $c_1 =\epsilon c_2$, $c_3=\epsilon c_2$ and $u_3-u_1=\epsilon^2 c_2$ and taking the limit $\epsilon \to 0$, we find that the above solution for the composite state becomes that for the  ${\langle 1,3 \rangle}$-lump as
\begin{align}
\phi_2^\pm=\frac{c_2}{\sd(u-u_1)},\quad \phi_3=\phi_1=0 \quad {\rm with} \quad {\bf m}={\bf m}^{\langle 1,3 \rangle }=(1,0,-1).\label{eq:singlelump2}
\end{align}
The existence of the continuous deformation from 
the ${\langle 1,2 \rangle}$, ${\langle 2,3 \rangle}$-lump composite to the ${\langle 1,3 \rangle}$-lump
clearly indicates that there are no topological obstacles in constructing the string junction illustrated in Fig.\,\ref{fig:JunctionImage}.

%%%%%%%%%%%%%%%%%%%%%%%%%%%%%%%%%%%%%
\section{Numerical solution of string junctions}\label{sec:junction}
In this section, we construct 
a numerical solution of string junctions 
%by the relaxation method
.
We go back to the original Lagrangian 
of the $F_2$ model of $N=3$ with the $S_3$ symmetry (the model with $r_1=r_2=r_3=f^2/2$)  
and turn on the %Faddeev-
Skyrme and potential terms.

%%%%%%%%%%%%%%%%%%%%%%
\subsection{Initial configuration and boundary condition}
\begin{figure}[ht]
\centering
\includegraphics[width=12cm]{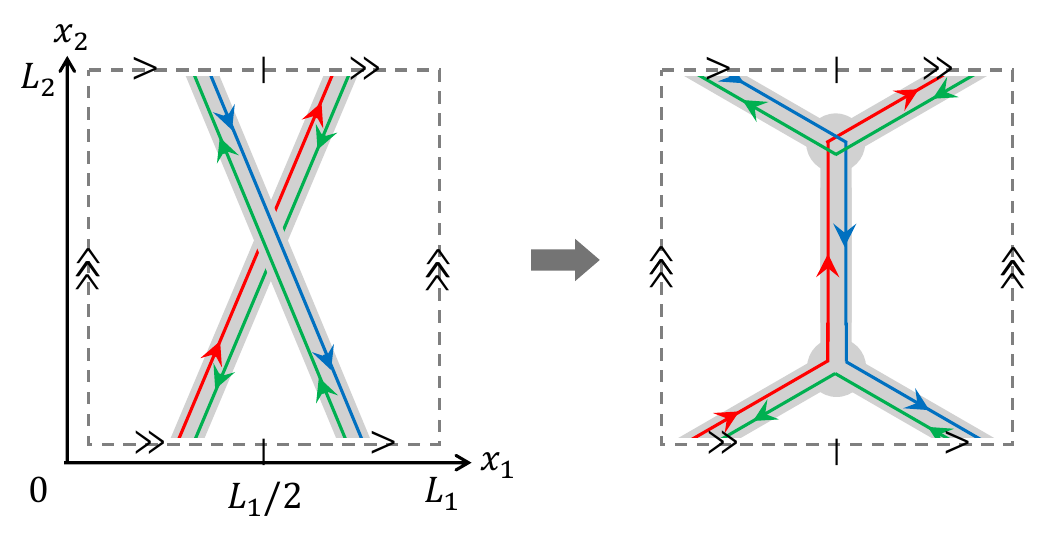}
\caption{ \label{fig:junctionforming} %Image for 
Schematic picture of a pair formation
of string junctions with an intersection of two strings through numerical optimization.%by using the relaxation method.
%({\color{red} Thank you, Amati-kun!   Could you also add arrows to the solid lines on the left panel? 
%Can you see from this diagram that the periodicity $f(x_1+L_1)\simeq f(x_1)$? })
}
\end{figure}

Our goal in this paper is to construct a numerical solution for a string junction illustrated in Fig.\,\ref{fig:JunctionImage}. 
To this end, 
we take the following strategy 
depicted in Fig.\,\ref{fig:junctionforming}. 
We take a torus $T^3$ as the base space to clarify the situation in numerical calculations. 
First, we prepare a pair of $\langle 1,2\rangle$ and $\langle 2,3 \rangle$-strings on the torus 
so that they form skew lines and nearly intersect each other, 
as illustrated in the left panel of Fig.\,\ref{fig:junctionforming}. 
Starting from this configuration as an initial condition, 
we perform a numerical optimization
to reduce the energy of the configuration. 
Then, the two strings will stick together in the middle, producing a $\langle 1,3\rangle$-string with a pair of the string junctions. 
As the energy decreases, the length of the $\langle 1,3\rangle$-string increases further until the angles between all the strings at the junction points become $2\pi/3$. 
The energy is expected to be minimized  
to reach a solution of the form illustrated in the right panel of Fig.\,\ref{fig:junctionforming}. 
The two types of junctions appearing here are related to each other by spatial rotation and complex conjugation.\footnote{
Here, a string is represented by parallel and opposite arrows, but their spatial 
ordering is merely for the convenience of the drawing and has no physical meaning. Therefore, the twisting of the arrows around the upper junction in the right panel of Fig.\,\ref{fig:junctionforming} has no physical meaning.}

More specifically, as the initial configuration, we choose the configuration given in Eq.\,\eqref{eq:compositestate} 
at arbitrary constant $x_2$ surfaces, set $c_1=c_3=1/2$ 
and give the $x_2$-dependence to the parameters $u_1, u_3$ as
\begin{align}
u_3&=u_3(x_2)\equiv \frac{1}{L_*} \left(\frac{L_1}4 +   \frac{L_1}{2L_2} x_2 + i\left(\frac{L_3}2+\delta\right)\right),\quad \nonumber\\
u_1&=u_1(x_2)\equiv \frac{1}{L_*} \left(\frac{3L_1}4 -  \frac{L_1}{2L_2} x_2 +i\left(\frac{L_3}2-\delta\right)\right),\quad  \label{eq:x2dependence}
\end{align}
where $\delta$ is a small constant 
introduced to avoid a singularity due to intersection of the strings.  
Here $u_3(x_2)$ and $u_1(x_2)$ satisfy  $u_3(x_2+L_2)=u_3(x_2)+K, u_1(x_2+L_2)=u_1(x_2)-K$.

On the base space $T^3$,
we impose the following twisted periodic boundary condition on the projection matrices $P_i$: 
\begin{align}
P_i(x_1+L_1,x_2,x_3)&= U_1^\dagger P_i(x_1,x_2,x_3) U_1,\quad U_1={\rm diag}(-1,1,1),\label{eq:Pperiod1}\\
P_i(x_1,x_2,x_3+L_3)&= U_3^\dagger P_i(x_1,x_2,x_3) U_3,\quad U_3={\rm diag}(1,1,-1),\label{eq:Pperiod2} \\ 
P_i\left(x_1+\frac{L_1}2,x_2+L_2,x_3\right)&=  P_i(x_1,x_2,x_3),
\label{eq:Pperiod3}
\end{align}
so that this periodicity is consistent with the initial condition given in Eqs.\,\eqref{eq:compositestate} and \eqref{eq:x2dependence}.
In Appendix \ref{sec:otherBC}, we explore alternative boundary conditions and demonstrate that they yield qualitatively similar results. 
Let us note here that the twist in this boundary condition is not essential to our goal of creating string junctions, but is only a technical expedient.

Under this setting, the lengths of the 
$\langle 1,2\rangle$ and $\langle 2,3\rangle$ strings 
in the fundamental domain should be 
$L_1/\sqrt{3}$ and that of the $\langle 1,3\rangle$-string becomes $L_2-L_1/(2\sqrt{3})$.

\subsection{Numerical results} \label{sec:NumericalResults}
Before considering a string junction, let us briefly describe the basic data of the component lump solution.
In the case with $1/g=\mu=0$, 
the configurations of single lump given in  Eqs.\,\eqref{eq:singlelump3}, \eqref{eq:singlelump1} and \eqref{eq:singlelump2} 
are still solutions even 
if $r_2$ is turned on.
There, as a result of scale invariance, 
a flat direction (zero mode) corresponding to the size moduli $|c_i|$ appears on the configuration space.
In the lattice calculations, however, due to the finite lattice spacing, 
this flat direction $|c_i|$ is slightly tilted toward the origin of $c_i$. Thus, during numerical optimization, 
the lump size shrinks toward zero and eventually the configurations break down when the lump %size is 
becomes smaller than the lattice spacing.
Therefore, we need to make both of the two parameters, $1/g$ and $\mu$, to be finite to numerically obtain stable lump solutions. %to stabilize the size in order to explicitly avoid this subtle problem on the numerical calculation.
With the two parameters, the lump size takes 
a value roughly estimated  
to be on the order of $1/\sqrt{\mu g}$
by the scaling argument.

A set of the values of the parameters %we choose 
for our numerical simulations is 
\begin{align}
  f^2=1,\quad  \frac1{g^2}=1, \quad \mu^2=1, \label{eq:parameters}
\end{align}
and the size of the torus is chosen as follows:
\begin{align}
    (L_1,L_2,L_3)=\frac{n}4 \times (8,7,5),\quad {\rm with~} n=3,4,5,6.
\end{align}
Here, $L_2/L_1=7/8$ is chosen to be a rational number close to $\sqrt{3}/2$, where the lengths of the three types of strings are approximately equal within the fundamental domain of $T^3$.

\begin{figure}[t]
\centering
\includegraphics[width=16cm]{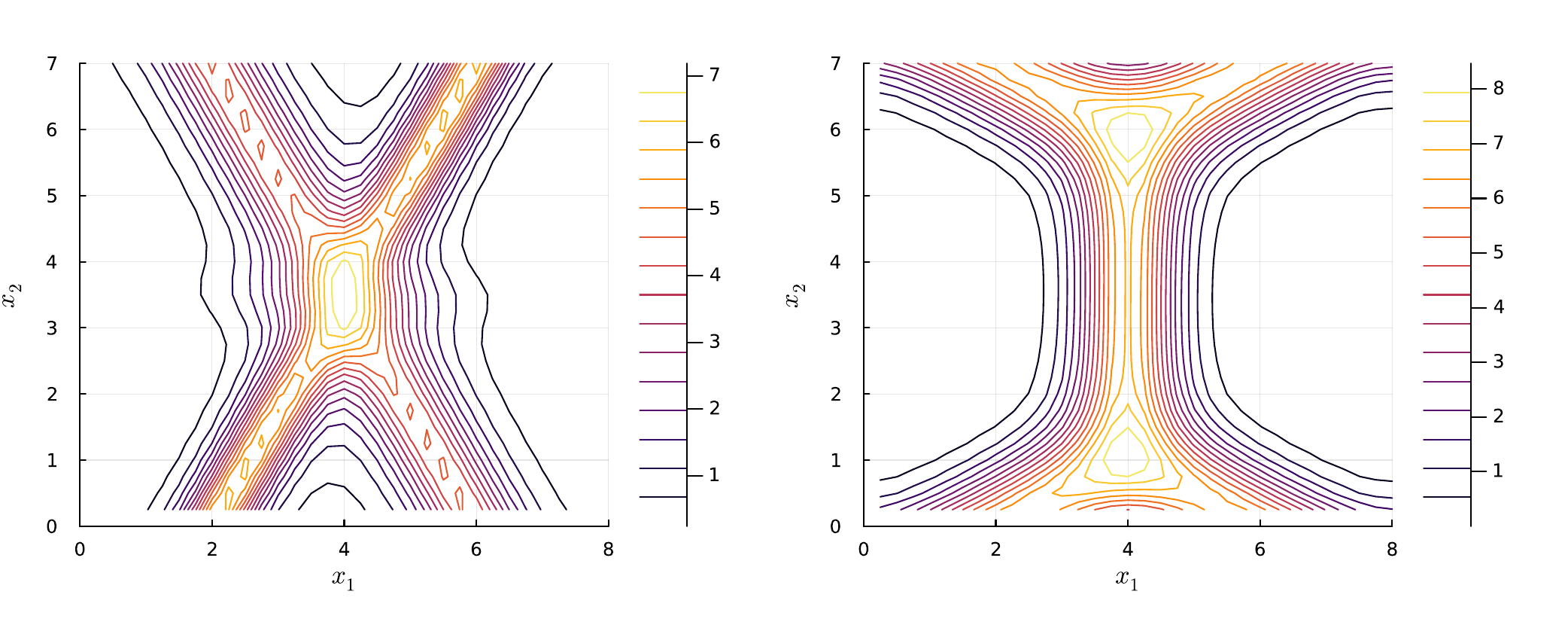}
\caption{ \label{fig:dencities}   
Energy densities at the $x_3=L_3/2$  cutting plane. The left panel is for the initial state, and the right panel is for the final state of the numerical simulation %relaxation
with $(L_1,L_2,L_3)=(8,7,5)$ and $a=1/4$.}
\end{figure}

\begin{figure}[t]
\centering
\includegraphics[width=15cm]{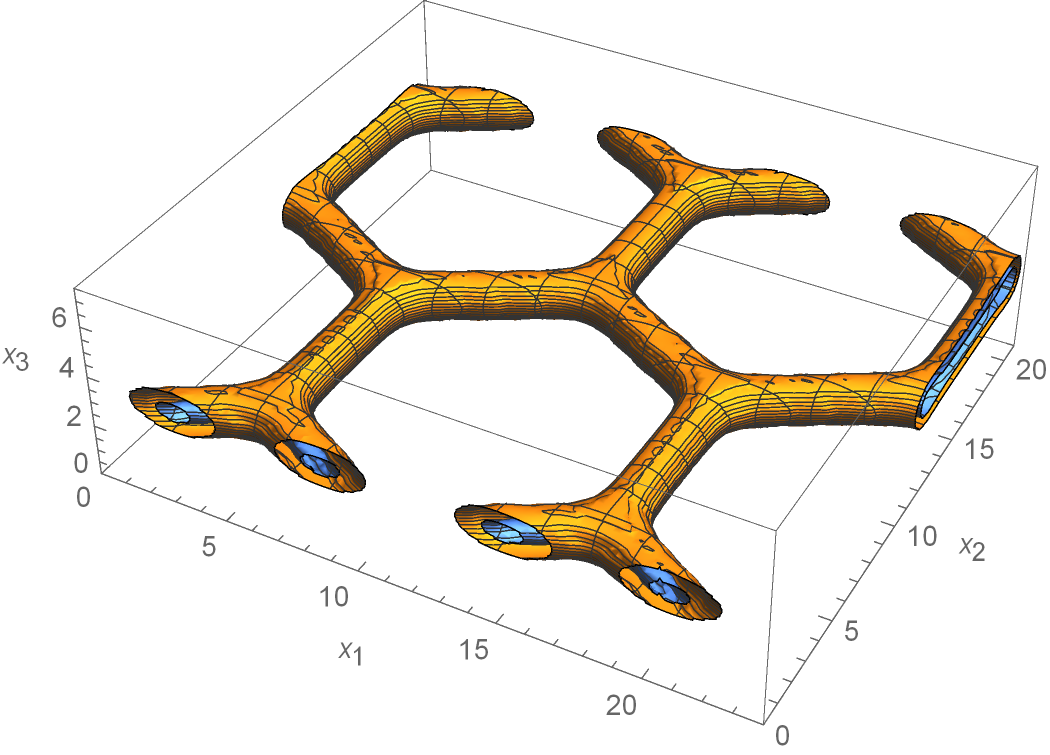}
\caption{ \label{fig:junctiongraph}   
The surface contour of $\rho=8$ (Orange) and $\rho=16$ (Blue) in the energy density $\rho$ for a net composed of the string junctions in the case with $(L_1,L_2,L_3)=(12,10.5,7.5)$ and $a=1/8$.
Four fundamental domains of the torus are shown.
}
\end{figure}
%With 
In these parameter settings, we numerically construct a configuration for a pair of the string junctions, 
which is our objective in this paper.
We conducted numerical simulations using the steepest descent method with a finite difference approximation where
the number of lattice points is $7560(=24\times 21\times 15)$ --  $483840(=96 \times 84\times 60)$, and the lattice spacing is taken as either $a=1/4$ or $a=1/8$.
A detailed description of our numerical method is given in Appendix.\,\ref{appendix:NumericalCalculation}.
%As illustrated in Fig.\,\ref{fig:junctionforming}, by applying the relaxation method to 
As the result of the numerical optimizations with the initial configuration given in Eqs.\,\eqref{eq:compositestate} and \eqref{eq:x2dependence}, 
we obtain the final converged configuration shown in 
Figs.\,\ref{fig:dencities} and \ref{fig:junctiongraph}, 
in which the energy density $\rho$ is depicted.

We also study the $L_i$ dependence of the total energy to remove a possible dependence on situational settings such as the twisted periodic boundary condition we have chosen for the numerical calculation.
 \begin{figure}[t]
\centering
\includegraphics[width=15cm]{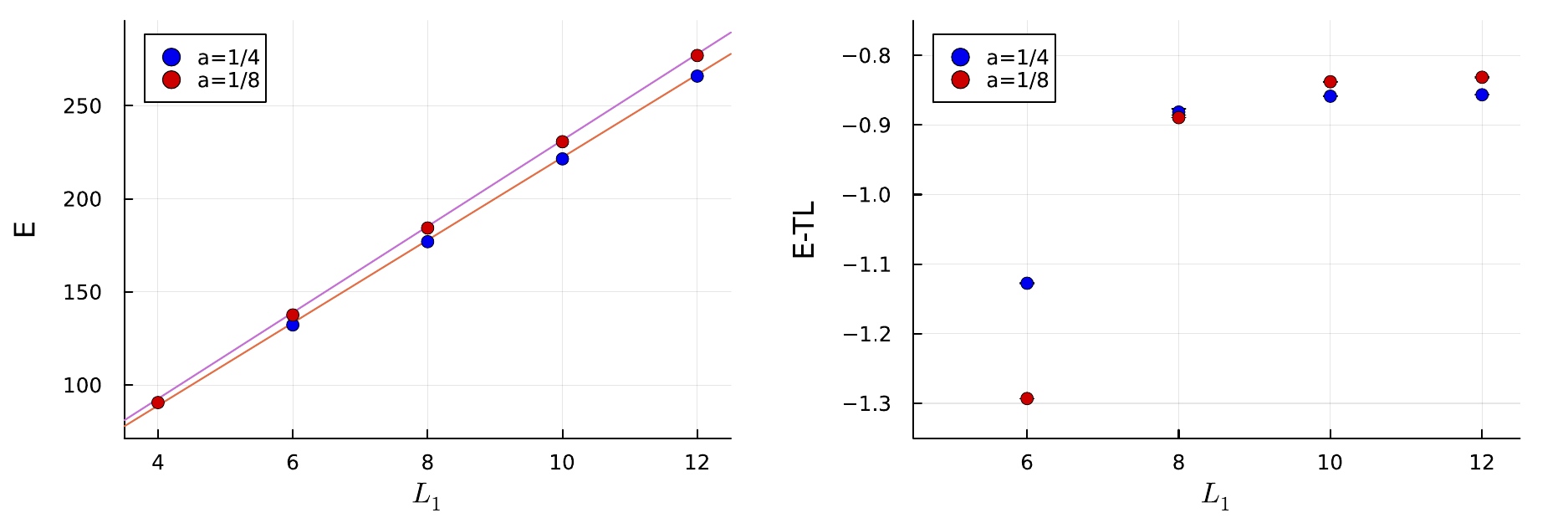}
\caption{$L_i$-dependence of total energy of the string junction for each fundamental domain of $T^3$, with the 
lattice spacing $a=1/4,1/8$ and keeping the ratio of the periods, $L_1:L_2:L_3=8:7:5$.}\label{fig:ELdependence}
\end{figure}
We plot numerical results for the total energy $E$ of the string junction in each fundamental domain $T^3$ by dots in Fig.\,\ref{fig:ELdependence}.
We assume that the energy $E$ behaves as
\begin{align}
    E =  T L +2M +{\cal O}(e^{-{\rm const.}\times L}),\quad {\rm with~}L=\frac{\sqrt{3}}2L_1+L_2,\label{eq:El}
\end{align}
where $T$ is the string tension and 
$M$ is a junction mass. 
In \cite{Komargodski:2024swh}, the authors have discussed that when $M$ is large and negative, quantum fluctuations around the string junction lead to an instability due to a tachyonic mode.
Note that there appears $2M$ because there is a pair of the junction and anti-junction in the fundamental domain.
Let us estimate the value of $M$ for our numerical solution. 
As represented by the solid lines in the left panel of Fig.\,\ref{fig:ELdependence},
the linear potentials $T L$ as contributions from the strings dominate the total energy and 
$T$ can be read as\footnote{
The values of $T$ are estimated by applying the same method as in Eq.\eqref{eq:modE} 
to the difference of the total energy $dE/dL_1$, by assuming the form \eqref{eq:El}. 
Although there can be a 
L\"uscher term proportional to $1/L$, 
it is irrelevant in our classical calculation.
The determination of the value of $M$ is strongly depends on how this $T$ is determined.}
\begin{align}
  %  T\times \left(\frac{\sqrt{3}}2L_1+L_2\right)  \quad {\rm with~} 
    T= \left\{ \begin{array}{cc}
       12.7706  &  {\rm for~} a=\frac14\\ \vspace{0.5mm}
       13.2993  &  {\rm for~} a=\frac18
    \end{array}\right. {~\rm with~} L_2=\frac 78 L_1.
\end{align}
Since the $a$-dependence of the deviation of the total energy from the continuum limit should be of order $a^2$, 
the total energy in the $a \rightarrow 0$ limit is  predictable by extrapolation.
For example, the continuum limit of the tension of the lump string with parameters given in Eq.\,\eqref{eq:parameters} is 
estimated to be $T \approx 13.5$.
In the right panel of Fig.\,\ref{fig:ELdependence}, we removed the contribution of the string tension from the total energy.
The value of $M$ in the $a \rightarrow 0$ reads roughly
\begin{align}
M \approx-0.4,  \label{eq:junctionmass}
\end{align}
which should be a quantity independent of the boundary conditions. See Appendix \ref{sec:otherBC} for cases with other boundary conditions.
\begin{figure}
\centering
\begin{minipage}{0.45\textwidth}
\includegraphics[width=6cm]{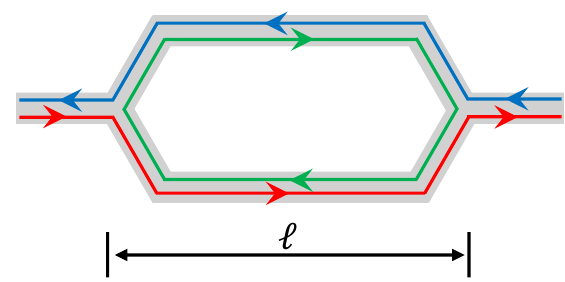}
\end{minipage}
\begin{minipage}{0.45\textwidth}
\includegraphics[width=6cm]{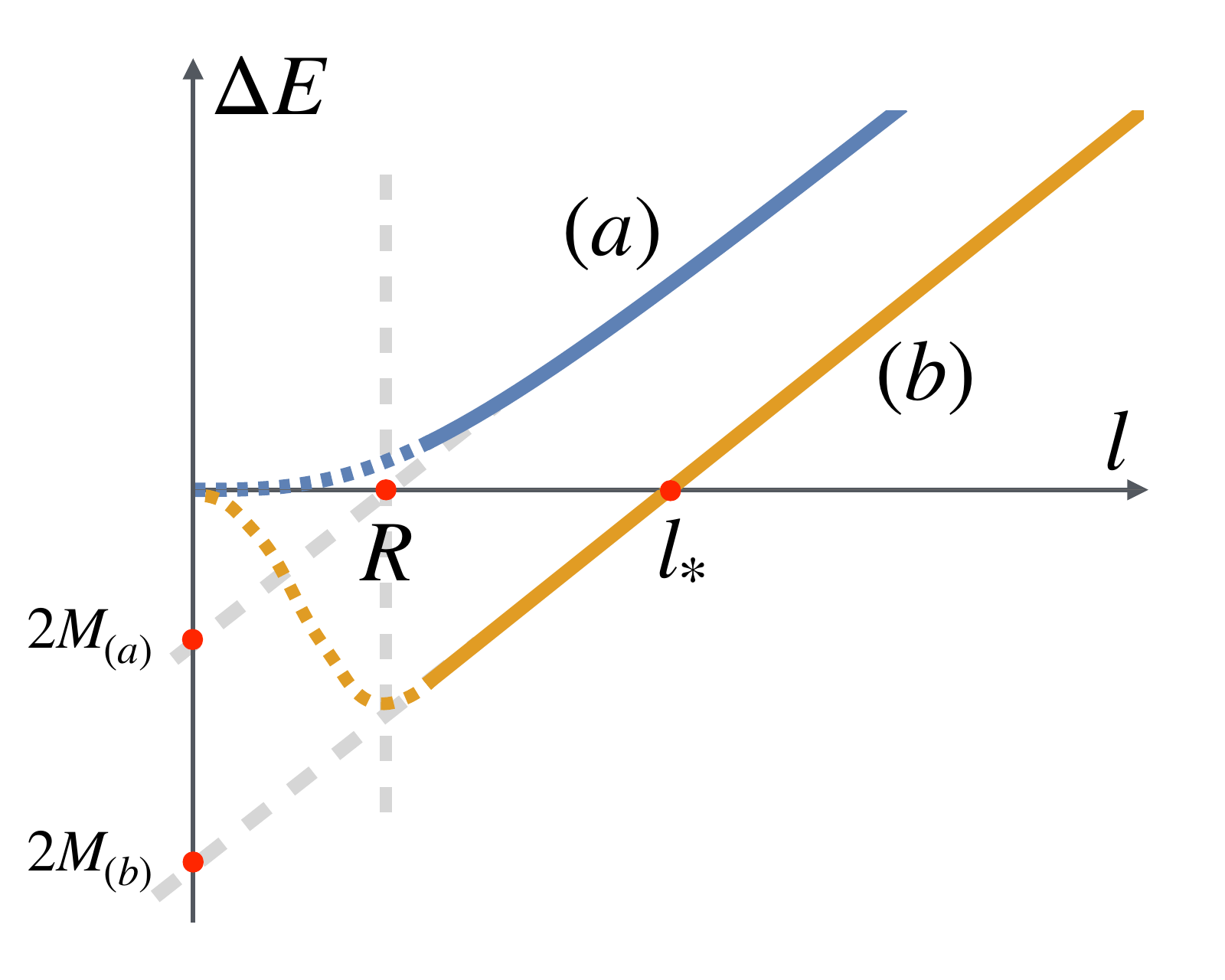}
\end{minipage}
\caption{Pair production of junctions and instability due to a large negative value of the junction mass: In the left panel, a pair of junction is created from a single string. Within the interval of length $l$ between the junctions, the string splits into two segments. The right panel shows the energy increment $\Delta E$ as a function of $l$ for two cases (a) $l_\ast \lesssim R$ and (b) $l_\ast \gg R$.
The cases of (a) and (b) correspond to stable and unstable 
junctions, respectively.
}\label{fig:Instability}
\end{figure}

Finally, let us check that our numerical solution is free from the possible instability caused by large negative values of $M$. 
Although it has been shown that the instability appears at the quantum level in \cite{Komargodski:2024swh}, even at the classical level, 
we can discuss the instability by considering a pair production of the string junctions as illustrated in the left panel of Fig.\,\ref{fig:Instability}.
The right panel of  Fig.\,\ref{fig:Instability} shows a sketch of the energy increment $\Delta E(l)$ from a single string configuration as a function of the distance between the junctions $l$.
For large $l$, the energy increment $\Delta E(l) \approx Tl + 2M$ is a linear function of slope $T$ and $y$-intercept $2M$. This naive estimate is valid as long as $l \gg R$, where $R$ is a typical scale of the string thickness or the interaction length between strings.
The lines (a) and (b) in Fig.\,\ref{fig:Instability} correspond to the two cases with $l_\ast \lesssim R$ and $l_\ast \gg R$, respectively. 
Here $l_\ast = -2M/T$ is the $x$-intercept, below which the naive estimate of the energy increment $\Delta E(l)\approx T l+2M$ becomes negative. 
The dotted curves extrapolate the lines 
(a) and (b) to be connected to the origin, 
that we do not fix in this paper.
As discussed below, the case (b) 
has an apparent instability.
Suppose that $M$ is negative and sufficiently large such that
\begin{align}
M \ll - \frac{TR}{2} ~~~ \mbox{or equivalently} ~~~ l_* \gg R.
\end{align}
Then, there is a region $l_\ast > l \gg R$ 
where we can rely on the naive estimate,
which suggest that the energy increment is negative $\Delta E < 0$, i.e. there exist configurations with lower energy than the single string configuration.  
Therefore, the string itself is unstable due to the pair production in the case (b).

Our result in Eq.\,\eqref{eq:junctionmass} 
corresponds to 
the case (a) consistent with global stability of the strings:
\begin{align}
     |M| \, \lesssim \, T R \, \approx \, T \times {\cal O}(1/\sqrt{g\mu}, f/\mu) \, = \, {\cal O}(10).
\end{align}

%%%%%%%%%%%%%%%%%%%
\section{Summary and  Discussion} \label{sec:summary}
We have numerically constructed a three-string junction of Y-shape in the $F_2$ %nonlinear 
sigma model with the four derivative %Faddeev-
Skyrme and potential terms 
with a typical set of parameters in the system.
The introduction of a potential term stabilizes the strings %thickness 
and thus the string junction 
is expected to always be stable in the system with an arbitrary set of parameters,
independent of the lattice spacing and the periods of the base space $T^3$.
To eliminate the effects caused by the finite periods of the torus and the finite lattice spacing, we have investigated the $a$-dependence and the $L_i$-dependence of the string junction.
In the large volume limit, 
it reduces to the junction in ${\mathbb R}^3$.
The stability of the junction has been also discussed.

The model taken in this paper has a dimensionless parameter $\mu/(gf^2)$ which is eventually taken to be 1 
for the numerical calculations. 
Furthermore, we can choose a different type of a four-derivative term and a potential term.
Details of the numerical results for the string junction will depend on such choices, however, it is expected that no drastic changes will occur. 

Here we address possible future directions and discussion. 
The original proposal by Faddeev and Niemi is 
that glueballs are described by Hopfions.
Thus far Hopfions in the $F_2$ Fadeev-Skyrme model 
was discussed only in Ref.\,\cite{Amari:2018gbq}.
Therefore, one of possible future directions 
is to construct Hopfions with junctions 
and to compare them with glueballs in 
$\SU(3)$ Yang-Mills theory.

Although we have concentrated on 
a three-string junction in the $F_2$ nonlinear sigma model, the model itself was defined for general flag manifold $F_N$.
In the case of the $F_3$ nonlinear sigma model,
we may show that a four-string junction of tetrapod-shape exists using the same manner we took in this paper.
%{\color{blue}[CAUTION: This tetrapod-shape junction is unstable in the almost generic cases including symmetric one! We can show that there is almost always a more stable pair of Y-shape junctions.] }

In this paper, 
we have used BPS strings in the $F_2$ sigma model just as an initial configuration for numerical %relaxations
simulations. 
Whether there is a model admitting 
a BPS string junction is actually an open question. 
The flag sigma model itself does not 
admit BPS string junctions. 
From a supersymmetric point of view, 
1/4 BPS equations for string junctions are proposed \cite{Naganuma:2001pu,Eto:2005sw,Gudnason:2020hcb} 
(see also Ref.\,\cite{Jackson:2006qc})
but no explicit solution is available.
Thus, we expect that  
there is a supersymmetric theory 
of any modification of the flag sigma model 
admitting a BPS string junction.

Cosmological consequences of confining strings as cosmic strings in pure Yang-Mills theory 
were studied in detail 
\cite{Yamada:2022imq}, 
and thus similar analysis could be 
performed for strings in the flag sigma models.
In particular, 
two strings with different topological charges do not reconnect in their collision. 
Fig.\,\ref{fig:junctionforming} in fact shows 
a production of a string stretched between two strings  after a collision of these strings with different topological charges.\footnote{
A similar phenomenon is known to happen 
for a collision of two non-commutative strings
\cite{Kobayashi:2008pk}. 
Unlike such a case, this occurs even 
Abelian strings in this paper.
} 
Therefore,  
a string network is formed in this model,  
giving an impact on cosmology.

%Note Added:
%While preparing the draft of this paper, 
%we are aware of the paper \cite{Komargodski:2024swh} discussing the string junction in QCD in ArXiv, 
%in which the stability condition of a 
%junction is discussed in terms of a junction mass. 
%It is an important future problem 
%that the junction mass that we obtained in 
%this paper 
%(the right figure of Fig.\,\ref{fig:ELdependence} in Appendix \ref{appendix:NumericalCalculation}) 
%is compared with their condition.

\section*{Acknowledgments}
YA would like to thank Nobuyuki Sawado and Keisuke Wada. This work is supported in part by JSPS KAKENHI [Grants No. JP23KJ1881 (YA), No. JP21K03558 (TF) and No. JP22H01221 (MN)], the WPI program ``Sustainability with Knotted Chiral Meta Matter (SKCM$^2$)'' at Hiroshima University.

%\newpage

\appendix
\section{Parametrization of the Lagrangian}
\label{sec:parametrization}
Faddeev and Niemi conjectured that the $F_{N-1}$ Faddeev-Skyrme model describes the low-energy limit of $\SU(N)$ Yang-Mills theory. In terms of their parametrization, the Lagrangian \eqref{eq:Lagrangian_P} can be written as
\begin{equation}
    -{\cal L}= \sum_{a=1}^{N-1}
    \left(\frac{f^2}{2}\tr\left[\partial_\mu n_a \partial^\mu n_a \right]+ \frac{1}{g^2} {\cal F}^a_{\mu\nu}{{\cal F}^a}^{\mu\nu} + \mu^2\tr\left[h_a(h_a-n_a)\right]  \right)  \ ,
    \label{eq:Lagrangian_n}
\end{equation}
where $h_a=\frac{1}{2}\lambda_{a(a+2)}$ is the basis of the Cartan subalgebra of $su(N)$ and the color direction fields $n_a$ are defined as
\begin{equation}
    n_a=U^\dagger h_aU \ ,
\end{equation}
with the $\SU(N)$ matrix $U$. The tensor ${\cal F}_{\mu\nu}^a$ are the coefficient of the Kirillov symplectic two-forms defined as
\begin{equation}
    {\cal F}_{\mu\nu}^a = -\frac{i}{2}\sum_{b=1}^{N-1}\tr\left[n_a[\partial_\mu n_b, \partial_\nu n_b] \right].
\end{equation}
In this appendix, we derive the Lagrangian \eqref{eq:Lagrangian_P} from Eq.\,\eqref{eq:Lagrangian_n}.

In addition to the basis of the Cartan subalgebra of $su(N)$, we introduce $h_N = \frac{1}{\sqrt{2N}}\mathbf{1}$.
The basis satisfies the orthonormal condition
\begin{equation}
   \tr[h_ah_b]=\frac{1}{2}\delta_{ab} \ .
   \label{eq:orthonormal_h}
\end{equation}
We decompose them in terms of the diagonal singleton matrices $p_i$ as
\begin{equation}
    h_a=\sum_{i=1}^N\nu_a^ip_i
    \label{eq:decomposition_h}
\end{equation}
with real constants $\nu_a^i$. 
The orthonormalzing condition \eqref{eq:orthonormal_h} implies that the vectors $(\nu_a^1,\nu_a^2,...,\nu_a^N)$ for $a=1,2,...,N$ form an orthogonal basis of $\mathbb{R}^N$ with the length $1/\sqrt{2}$. Therefore, the matrix $\nu$ is orthogonal, and we find
\begin{equation}
\sum_{a=1}^N\nu_a^i\nu_a^j=\frac{\delta_{ij}}{2},
\qquad
\sum_{i=1}^N\nu_a^i\nu_b^i=\frac{\delta_{ab}}{2} \ .
\end{equation}

From Eq.\,\eqref{eq:decomposition_h}, we can decompose the color direction fields $n_a$ in terms of the projectors $P_i$ as
\begin{equation}
    n_a=\sum_{i=1}^N\nu_a^iP_i \ .
\end{equation}
Substituting it into the first term in Eq.\,\eqref{eq:Lagrangian_n}, we obtain
\begin{align}
    \sum_{a=1}^{N-1}\tr\left[\partial_\mu n_a \partial^\mu n_a \right] 
    =&\sum_{a=1}^{N}\tr\left[\partial_\mu n_a \partial^\mu n_a \right] 
    \notag\\
    =&\sum_{a=1}^{N}\nu_a^i\nu_a^j\tr\left[\partial_\mu P_i \partial^\mu P_j \right] 
    \notag\\
    =&\frac{1}{2}\sum_{i=1}^{N}\tr\left[\partial_\mu P_i \partial^\mu P_i \right] 
    \label{eq:D-term_nP}
\end{align}
where we have used $\partial_\mu n_N=0$ because of $n_N=h_N$. Similarly, we can write
\begin{align}
    {\cal F}_{\mu\nu}^a=&-\frac{i}{2}\sum_{b=1}^{N}\tr\left[ n_a[\partial_\mu n_b, \partial_\nu n_b]\right]
    \notag\\
    =&-\frac{i}{2}\sum_{b=1}^{N}\sum_{j,k,l=1}^N
    \nu_a^j\nu_b^k\nu_b^l
    \tr\left[ P_j[\partial_\mu P_k, \partial_\nu P_l]\right]
    \notag\\
    =&-\frac{i}4\sum_{j,k=1}^N\nu_a^j
    \tr\left[ P_j[\partial_\mu P_k, \partial_\nu P_k]\right] \ .
\end{align}
Now, we have an identity for $i\neq j, j\neq k, k\neq l$ of the form
\begin{equation}
    \tr[P_i\partial_\mu P_j \partial_\nu P_k]=\tr[p_i[p_j,U\partial_\mu U^\dagger][p_k,U\partial_\nu U^\dagger]]=0 \ ,
\end{equation}
which is implied by $p_ip_j=\delta_{ij}p_j$. Using the identity, we obtain
\begin{align}
    {\cal F}_{\mu\nu}^a
    &=-\frac{i}{4}\sum_{j=1}^N\nu_a^j\left\{ \tr[P_j[\partial_\mu P_j,\partial_\nu P_j]] 
    +\sum_{k\neq j}\tr[P_j[\partial_\mu P_k,\partial_\nu P_k]]\right\}
    \notag\\
    &=-\frac{i}{4}\sum_{j=1}^N\nu_a^j\left\{ \tr[P_j[\partial_\mu P_j,\partial_\nu P_j]] 
    +\tr[P_j[\sum_{k\neq j}\partial_\mu P_k,\sum_{l\neq j}\partial_\nu P_l]]\right\}
    \notag\\
    &=-\frac{i}{2}\sum_{j=1}^N\nu_a^j\tr[P_j[\partial_\mu P_j,\partial_\nu P_j]] 
    \notag\\
    &=\frac{1}{2}\sum_{j=1}^N\nu_a^jF_{\mu\nu}^j
\end{align}
where we have used $\displaystyle{\sum_{i=1}^N}\partial_\mu P_j=0$.
Therefore, we find that the %Faddeev-
Skyrme term, the second term in Eq.\,\eqref{eq:Lagrangian_n}, can be cast into the form
\begin{align}
    \sum_{a=1}^{N-1}{\cal F}_{\mu\nu}^a{{\cal F}^a}^{\mu\nu}
    &=\sum_{a=1}^{N}{\cal F}_{\mu\nu}^a{{\cal F}^a}^{\mu\nu}
    \notag\\
    &=\frac{1}{4}\sum_{a=1}^{N}\sum_{i,j=1}^N\nu_a^i\nu_a^jF_{\mu\nu}^i{F^j}^{\mu\nu}
    \notag\\
    &=\frac{1}{8}\sum_{i=1}F_{\mu\nu}^i{F^i}^{\mu\nu} \ .
    \label{eq:Skyrme-term_nP}
\end{align}
In addition, the potential term can be written as
\begin{align}
    \sum_{a=1}^{N-1}\tr[h_a(h_a-n_a)]
    &=\sum_{a=1}^{N}\tr[h_a(h_a-n_a)]
    \notag\\
    &=\sum_{a=1}^{N}\sum_{i,j=1}^{N}\nu_a^i\nu_a^j\tr[p_i(p_j-P_j)]
    \notag\\
    &=\frac{1}{2}\sum_{i=1}^N\tr[p_i(p_i-P_i)]
    \notag\\
    &=\frac{1}{4}\sum_{i=1}^N\tr[(p_i-P_i)^2].
    \label{eq:potential_nP}
\end{align}
Summarizing the results in Eqs.\,\eqref{eq:D-term_nP},\eqref{eq:Skyrme-term_nP}, and \eqref{eq:potential_nP}, we find that the Lagrangian in Eq.\,\eqref{eq:Lagrangian_n} is equivalant to the one in Eq.\,\eqref{eq:Lagrangian_P} which we have studied in this paper.
%%%%%%%%%%%
\section{\texorpdfstring{$F_{N-1}$}{F(N-1)} sigma model with general coefficients }\label{sec:generalFN-1}
In this appendix we introduce the most general form of the $F_{N-1}$ sigma model.
The model %kinetic term 
can maximally possess %be the most generally extended with
 $N(N-1)/2$ parameters as %to 
\begin{align}
-{\cal L}_{\sigma\text{-model}}%K
=-\frac{1}2 \sum_{j=2}^{N}\sum_{i=1}^{j-1}f_{ij}\tr[\p_\mu P_i \p^\mu P_j] .
\label{eq:generalK}   
\end{align}
This general %ized
model returns to the original one by setting $f_{ij}=f^2$ for all $i,j$.
Here the coefficient $f_{ij}=f_{ji}$ must be positive definite. This can be confirmed as follows.
Since $F_{N-1}$ is a homogeneous complex manifold, we only need to examine the neighborhood of the origin.
Substituting  an unitary matrix $U\approx {\bf 1}+X$ with an anti-Hermitian matrix $X$ to the projection matrices 
$P_i=U^\dagger p_i U$ and taking quadratic terms in the Lagrangian, we find that
\begin{align}
  -{\cal L}_{\sigma\text{-model}}%K
  \approx  \sum_{j=2}^{N}\sum_{i=1}^{j-1}f_{ij} |\p_\mu X^i{}_j|^2,  \quad 
\end{align}
which shows that $f_{ij}$ must be positive definite; if it meets this condition, any value is acceptable.
Here, an $\langle i, j \rangle$-string remains to be a solution under this deformation 
to the general %ized
$F_{N-1}$ sigma model, because it is a solution of the $F_1\simeq \mathbb CP^1$ sigma model embedded 
into the $F_{N-1}$ one.
Note that %although 
the solution might be unstable as a saddle point.

In the absence of the %Faddeev-
Skyrme 
and potential terms ($1/g^2=\mu^2$=0),
the $\langle i, j \rangle$-string is a BPS solution 
whose tension is given by 
\begin{align}
    2\pi f_{ij}   
\end{align}
if and only if $f_{ij}\le f_{ik}+f_{kj}$ for all $k \not=i,j$ \cite{Fujimori:2023wkd}.
If $f_{ij}< f_{ik}+f_{kj}$, an $\langle i,j\rangle$-string is energetically more stable than a composite state of 
$\langle i,k\rangle$- and 
$\langle k,j\rangle$-strings. 
If $f_{ij}> f_{ik}+f_{kj}$, the $\langle i,j\rangle$-string is unstable and will separate into $\langle i,k\rangle$- and 
$\langle k,j\rangle$-strings. In the case with $f_{ij}= f_{ik}+f_{kj}$, there is no interaction between the $\langle i,k\rangle$- and 
$\langle k,j\rangle$-strings, and actually a composite state of them at any relative distance is a BPS state, 
which is exactly what is discussed in Sec \ref{sec:BPS-lump}.

By setting $f_{ij}=r_i+r_j$ with introducing $N$ parameters $\{r_i | i=1,2,\cdots,N\}$, the above general model reduces to 
a sum of $N$ copies of the $\mathbb CP^{N-1}$ sigma model of which Lagrangian is given by the simple  extension of  Eq.\,\eqref{eq:CPLsum},
 where a set $\{r_i\}$ must satisfy 
\begin{align}
    r_i+r_j >0   \quad {\rm for~} i\not = j.
\end{align}
For cases with $N\ge 4$, the above model contains $S_N$-symmetric point but covers only an $N$-dimensional subspace of the parameter space for the general case.
The $N=3$ case is special where parameter space of Eq.\,\eqref{eq:generalK} and 
one in Eq.\,\eqref{eq:CPLsum} are equivalent since
\begin{align}
&f_{ij}=r_i+r_j  \nonumber \\
\Leftrightarrow \quad &r_1=\frac12(f_{12}+f_{13}-f_{23}), \quad r_2=\frac12(f_{12}+f_{23}-f_{13}),\quad r_3=\frac12(f_{13}+f_{23}-f_{12}),
\end{align}
and thus the model given in Eq.\,\eqref{eq:CPLsum} is the most general. 

%%%%%%%%%%%%%%%%%%%%%%%%%%%
\section{Numerical calculation} \label{appendix:NumericalCalculation}
In this appendix, we describe some details of our numerical calculations.

\subsection{\texorpdfstring{$F_{N-1}$}{F(N-1)} on the lattice}
Let us discretize  the system by taking a $d$-dimensional Euclidean lattice $\Gamma$ as the base space, where the action is a function whose variables are a set of unitary matrices $U_{\vec x}\in \U(N)$ defined at each point $\vec x \in \Gamma$ as,
\begin{align}
    S^{\rm o}=a^d\sum_{\vec x \in \Gamma}\sum_{i=1}^{N}\left[ 
    \frac{f^2}{2a^2}  \sum_{j =1, (j\not=i)}^{N}\sum_{\mu=1}^d \tr[P_i^{\vec x+\vec \mu} P_j^{\vec x}] +\frac{1}{8g^2}\sum_{\mu,\nu=1}^d(F_{\mu\nu}^{i,\vec x})^2 +\frac{\mu^2}{4}\tr[(P_i^{\vec x}-p_i)^2] \right]
\end{align}
with the projection matrices $P_i^{\vec x}= U_{\vec x}^\dagger p_i U_{\vec x}$, 
the lattice spacing $a$ and $\vec \mu$  defined to point to an adjacent lattice point as $(\vec \mu)^\nu=a \delta^\nu_\mu$.
The field strength $f_{\mu\nu}^{i,\vec x}$ is defined as  
\begin{align}
   F_{\mu\nu}^{i,\vec x} \equiv  \frac 1{a^2} {\rm Im}\, \tr[ P_i^{\vec x}P_i^{\vec x+\vec \mu}P_i^{\vec x+\vec \mu+\vec \nu}P_i^{\vec x+\vec \nu}]
   =F_{\mu\nu}^i\left(\vec x+\frac{\vec \mu+\vec \nu}2\right)+{\cal O}(a^2).
\end{align}
Note that, to define an energy density  $\rho_{\vec x}$ at $\vec x\in \Gamma$ used in Figs.\,\ref{fig:dencities} and \ref{fig:junctiongraph},
the values at the relevant adjacent lattice points need to be averaged 
since the difference and  the field strength on $\Gamma$ are defined on the links and the plaquettes of $\Gamma$, respectively.

In order to deal with the variation  with respect to the unitary matrices, it is convenient to introduce Lagrange multipliers  $\lambda_{\vec x}$ which are $N$-th order Hermitian matrices defined at each $\vec x \in \Gamma$ and to add terms to the original action $S^{\rm o}$ as follows:
\begin{align}
    S=S^{\rm o}-a^d\sum_{\vec x \in \Gamma} \tr[\lambda_{\vec x}(U_{\vec x}^\dagger U_{\vec x} -{\bf 1}_{N})],\quad {\rm with~} \lambda_{\vec x}^\dagger = \lambda_{\vec x},
\end{align}
where the matrix $U_{\vec x}$ is not restricted to unitary.
Here, thanks to the $\U(1)^{N}$ gauge symmetry, $U_{\vec x}$ does not appear explicitly in $S^{\rm o}$ and 
the function $S^{\rm o}$ is written in terms of the matrices $P_i^{\vec x}$.   
Therefore, we can define a variation of $S^{\rm o}$ with respect to $P_i^{\vec x}$ as
\begin{align}
Q_i^{\vec x} \equiv \frac{\delta' S^{\rm o}}{\delta' P_i^{\vec x}}
=(Q_i^{\vec x})^\dagger,
\quad \Leftrightarrow 
\quad 
\delta S^{\rm o}(\{P_i^{\vec x}\})\equiv a^d\sum_{\vec x\in \Gamma} \sum_{i=1}^{N} \tr[\delta P_i^{\vec x} Q_i^{\vec x}],
\end{align}
where each $P_i^{\vec x}$ is treated as if it were an arbitrary Hermitian matrix unrelated to each other, and $\delta'$ means a variation under that manner.\footnote{
In other words, $Q_i^{\vec x}$ is defined by extending the function $S^{\rm o}(\{P_i^{\vec x}\})$ by extrapolation outside the domain of definition of $P_i^{\vec x}$.  
Strictly speaking this extension and thus the definition of $Q_i^{\vec x}$ are not unique, 
 but there should be no problem with this definition 
 since the introduction of $Q_i^{\vec x}$ is only for notational simplicity. 
 %{\color{red} To colleagues, see Appendix.\ref{sec:ToColleagues} for details}
 }

Under the above preparations, we can derive a variation of the action as
\begin{align}
    \delta S=a^d\sum_{\vec x\in \Gamma}
    \tr\left[i U_{\vec x}^{-1} \delta U_{\vec x} {\cal H}_{\vec x} +{\rm h.c.}\right]
\end{align}
with the matrix ${\cal H}_{\vec x}$ defined by
\begin{align}
  {\cal H}_{\vec x} \equiv -i\left(\sum_{i=1}^{N}Q_i^{\vec x}P_i^{\vec x} -\lambda_{\vec x}\right)= \frac{i}2 \sum_{i=1}^{N}\left[P_i^{\vec x},Q_i^{\vec x}\right], 
\end{align}
where we require that ${\cal H}_{\vec x}$ is Hermitian, ${\cal H}_{\vec x}={\cal H}_{\vec x}^\dagger$. This  requirement gives an equation 
solved for the Lagrange multiplier as,
\begin{align}
    \lambda_{\vec x}=\frac12 \sum_{i=1}^{N}\left\{P_i^{\vec x},  Q_i^{\vec x}\right\} 
    \quad \left(= (\lambda_{\vec x})^\dagger\right).
\end{align}
Since there 
are $N$ identities on ${\cal H}_{\vec x}$, $\tr[P_i^{\vec x}{\cal H}_{\vec x} ]=0$  under the unitary condition $U_{\vec x}^\dagger U_{\vec x}={\bf 1}$,
the number of independent equations in ${\cal H}_{\vec x}=0$ is $N(N-1)$ which is just the dimension of $F_{N-1}$.
Therefore, ${\cal H}_{\vec x}=0$ is nothing more than the equation of motion in this system.

\subsection{Numerical method}
To obtain numerical solutions to ${\cal H}_{\vec x}=0$, we apply a gradient method, %relaxation 
steepest descent, to this system.
We set each step of the numerical calculation %relaxation method 
on the set of the unitary matrices $\{U_{\vec x}|\vec x \in \Gamma \}$ as
\begin{align}
 U_{\vec x} \quad \to \quad    U^{(\alpha)}_{\vec x} =U_{\vec x} e^{i \alpha {\cal H}_{\vec x}},\qquad   
   \left( P_i^{\vec x,(\alpha)}=e^{-i \alpha {\cal H}_{\vec x}} P_i^{\vec x} e^{i \alpha {\cal H}_{\vec x}}\right),
\end{align}
with an appropriate step size $\alpha \in \mathbb R_{>0}$.
If we choose a sufficiently small $\alpha$, this method ensures that the total energy is always decreasing at each step, because
\begin{align}
   \lim_{\alpha\to +0} 
   \frac1\alpha \left[S(\{P_i^{\vec x,(\alpha)}\}) -S(\{P_i^{\vec x}\})\right]=-2a^d\sum_{\vec x \in \Gamma} \tr {\cal H}_{\vec x}^2 \quad \le 0.
\end{align}
Note that, since ${\cal H}_{\vec x}$ is Hermitian,  each step of %the relaxation method 
iterations will automatically keep the unitary condition 
$U^\dagger U={\bf 1}$, if the initial condition satisfies it.

After enough iterations of the steps in 
%the relaxation method
a numerical simulation, the deviation from the solution of the fields $\phi_i^{\vec x}$ decreases exponentially as 
$\delta \phi_i^{\vec x} \approx \varphi_i^{\vec x} e^{-\Delta t}$,
where $\varphi_i^{\vec x}$ is the lightest massive mode around the solution and  $\Delta$ is a certain positive real number related to the mass gap and  
$t$ is the relaxation time as the accumulation of $\alpha$ in each step up to that point.
Under the %relaxation
optimization, therefore, that of the total energy $E$ behaves as $\delta E\propto e^{-2\Delta t}$ whereas the other observed quantities $O_a$  behave
as $\delta O_a \propto e^{-\Delta t}$.
Using this knowledge, a faster converging sequence of numbers $\{\hat E_n\}$ can be constructed from the sequence  $\{E_n\}$ obtained by the gradient %relaxation
method as follows,
\begin{align}
    \hat E_n \equiv \frac{E_{n}E_{n-2}-E_{n-1}^2}{E_{n}+E_{n-2}-2E_{n-1}}, \label{eq:modE}
\end{align}
and then the calculation error  can be roughly estimated as $|\hat E_n-E_n|$
when the calculation is terminated at a certain $n$.\footnote{Due to truncation errors from taking the difference, the calculation of $\hat E_n$ requires higher computational accuracy than that of $E_n$.
Therefore, if the required accuracy for $\hat E_n$ exceed the calculation accuracy, then this estimation does not work well.}

In Sec.\,\ref{sec:NumericalResults}, 
we applied the gradient %relaxation
method explained above to %the numerical calculation for constructing 
the construction of the string junctions.
The boundary conditions we have adopted here are technical and not essential to the construction of the string junction.
 In order to estimate and remove the effect of the periodic boundary conditions, 
 we plot the $L_i$-dependence of the total energy in Fig.\,\ref{fig:ELdependence} with keeping the ratio $L_1:L_2:L_3=8:7:5$.
In those numerical calculations the termination conditions were set as follows 
\begin{align}
 \forall \vec x\in \Gamma: \qquad\sqrt{\tr[{\cal H}_{\vec x}^2]} \le {\cal O}(10^{-4})  \quad {\rm or~} {\cal O}(10^{-3})   
\end{align}
and then  calculation errors in the total energy are estimated to be  $|\hat E_n-E_n| < 5\times 10^{-4}$ 
which can be omitted in Fig.\,\ref{fig:ELdependence}.  
It can be confirmed that these numerical solutions satisfy at least $\sqrt{\tr[(U^\dagger_{\vec x} U_{\vec x}-{\bf 1})^2 ]}\le {\cal O}(10^{-11})$, and as explained above, the unitary condition is almost preserved.

\section{Alternative boundary conditions}\label{sec:otherBC}
In the main text, we imposed specific boundary conditions given by
\eqref{eq:Pperiod1}-\eqref{eq:Pperiod3}. 
Here, we check that the configurations near junction points are independent of the boundary conditions by numerically solving the equations of motion under alternative boundary conditions.

\subsection{Untwisted torus}
The contribution of the string tension dominates the total energy of string junction configurations. Thus, given the fundamental region of the torus $T^3$, the length of each of the strings is uniquely determined to minimize its energy, and hence the junction points are also uniquely determined, except for the translational degrees of freedom. 
In particular, if the tensions of all strings are equal, as in the main text, the angles between strings at the junction should be $2\pi/3$ due to the balance of tensions.

As an example of an untwisted torus, we impose the following boundary conditions:
\begin{align}
P_i(x_1+L_1,x_2,x_3)&= U_1^\dagger P_i(x_1,x_2,x_3) U_1,\quad U_1={\rm diag}(-1,1,1),\label{eq:untwistedPBC11}\\
P_i(x_1,x_2,x_3+L_3)&= U_3^\dagger P_i(x_1,x_2,x_3) U_3,\quad U_3={\rm diag}(1,1,-1),\label{eq:untwistedPBC12}\\ 
P_i\left(x_1,x_2+L_2,x_3\right)&=  P_i(x_1,x_2,x_3),
\label{eq:untwistedPBC13}
\end{align}
where only the third condition \eqref{eq:Pperiod3} has been replaced.
As an initial configuration for the gradient method, we prepare two orthogonal strings that are parallel to the $x_1$ and $x_2$ axes as follows.  
We set $L_1=L_2$ for simplicity and adopt the functional forms given in Eq.\,\eqref{eq:compositestate} for $(\phi_1,\phi_2,\phi_3)$ as in the main text,
but use the different parametrization for $(u,u_1,u_3)$: 
\begin{align}
    u&=\frac1{L_*}\left\{\frac12(x_1+x_2)+i  x_3\right\},\nn\\
    u_3&=\frac1{L_*}\left\{\frac12(x_2-x_1+L_1)+i \left(\frac{L_3}2 +\delta\right)\right\}, \nn\\
u_1&=\frac1{L_*}\left\{-\frac12(x_2-x_1-L_2)+i\left(\frac{L_3}2 -\delta\right)\right\}.
\end{align}
With this parametrization, $\phi_3$ is meromorphic with respect to $x_1+ix_3$ and represents a straight $\langle 1,2\rangle$-string parallel to the $x_2$-axis and $\phi_1$ is meromorphic with respect to $x_2+ix_3$ and represents a straight $\langle 2,3\rangle$-string parallel to the $x_1$-axis.

\begin{figure}[h]
\centering
\includegraphics[width=16cm]{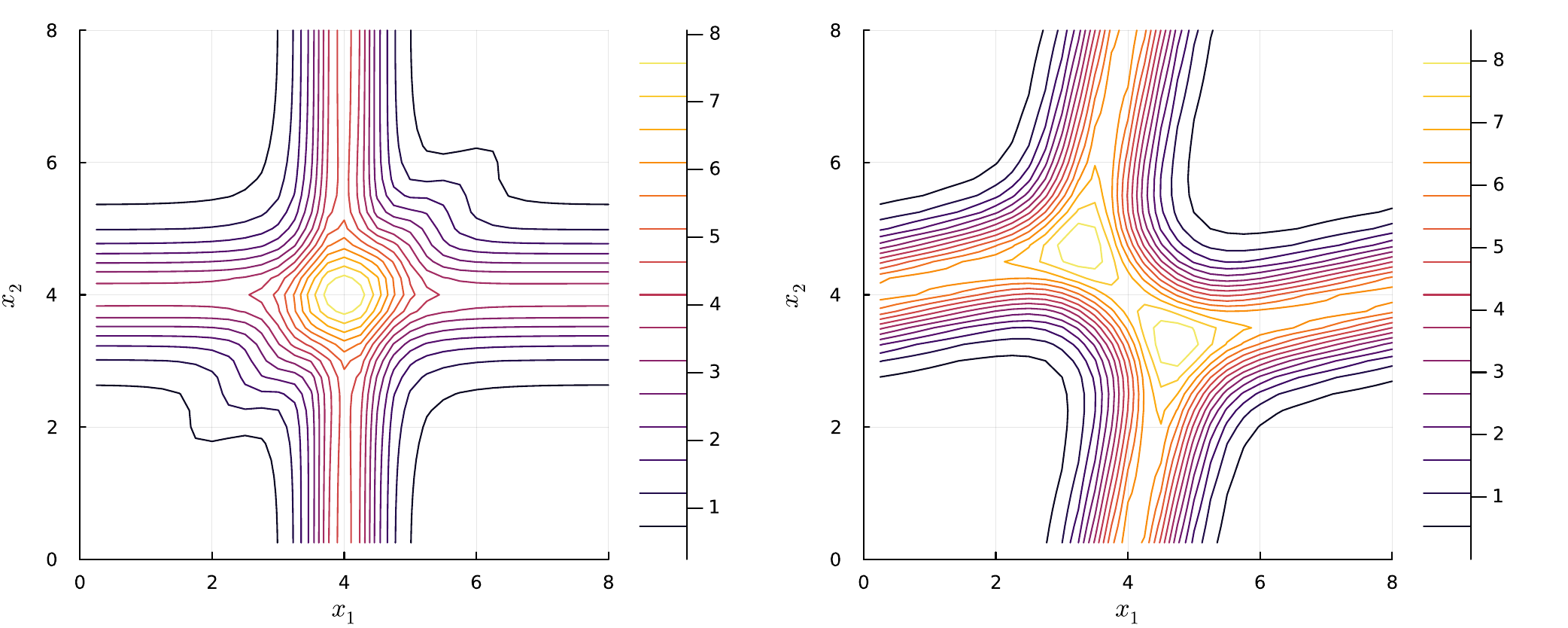}
\caption{ \label{fig:dencitiesInitailCross}   
Energy densities at the $x_3=L_3/2$ cutting plane. The left panel is for the initial state, and the right panel is for the final state of the numerical simulation %relaxation
with $(L_1,L_2,L_3)=(8,8,5)$ and $a=1/4$.}
\end{figure}

Using this initial configuration, we performed the numerical calculations under the untwisted boundary conditions \eqref{eq:untwistedPBC11}-\eqref{eq:untwistedPBC13}. Energy density for the resulting configuration with $L_1=8$ is shown in Fig.\ref{fig:dencitiesInitailCross}, where a pair of string junctions and a $\langle 1,3\rangle$-string stretched between them appear as in the case discussed in the main text. 
From the balance of string tensions, 
the string lengths in the fundamental domain  can be determined as $\sqrt{2/3} \, L_1$ for the $\langle 1,2\rangle$- and $\langle 2,3\rangle$-strings and   $(1/\sqrt{2}-1/\sqrt{6}) L_1$ for the $\langle 1,3\rangle$-string.
We can confirm for $L_1=8$ that the contribution from the string tension is dominant in the total energy $E$ in the fundamental domain: 
\begin{align}
    \left|E-T\times\frac{1+\sqrt{3}}{\sqrt{2}} L_1\right|= \left|195.894 - 12.8\times\frac{1+\sqrt{3}}{\sqrt{2}}\times 8\right|\ll E .
\end{align}
%This absolute value of this difference is considerably larger than that ($\approx -0.9$) for $L_1=8$ in the main text. This may be due to the fact that the shortest distance between the two junction points is smaller than that for the same $L_1$ in the main text.

In this configuration, the distance between the junction points is shorter than that in Fig.\,\ref{fig:dencities} and hence the interaction between the junction points is not negligible.
To minimize finite area effects, 
a twisted torus where all strings length are equal, as adopted (approximately) in the main text, is more suitable as the base space.

\subsection{Periodic boundary condition without flavor symmetry twist}
In the main text, we imposed the periodic boundary conditions with a flavor symmetry twist as given in Eqs.\,\eqref{eq:Pperiod1}-\eqref{eq:Pperiod3}.
Here, let us replace them with the following periodic boundary conditions without the flavor symmetry twist:
\begin{align}
P_i(x_1+L_1,x_2,x_3)&= P_i(x_1,x_2,x_3),\label{eq:untwistedPBC21}\\
P_i(x_1,x_2,x_3+L_3)&= P_i(x_1,x_2,x_3),\label{eq:untwistedPBC22} \\ 
P_i\left(x_1+L_1/2,x_2+L_2,x_3\right)&=  P_i(x_1,x_2,x_3).\label{eq:untwistedPBC23}
\end{align}
To obtain solutions satisfying the above boundary conditions by the gradient method,  
we impose the same boundary conditions on the initial configuration.
However, as mentioned in Sec.\,\ref{sec:BPS}, configurations $(\phi_1,\phi_2,\phi_3)$ that are meromorphic in $x_1+ix_3$ cannot satisfy these untwisted boundary conditions. To prepare an initial configuration satisfying the above boundary conditions, it is necessary to modify the initial configuration given in  Eqs.\,\eqref{eq:compositestate} and \eqref{eq:x2dependence} by relaxing the meromorphic constraint on $(\phi_1,\phi_2,\phi_3)$. We can show that a suitable initial configuration can be obtained by replacing $c_3=c_1=1/2$ as
\begin{align}
c_3&=\frac12 \cos\left(\frac{\pi}{2K}{\rm Re}(u-u_3)\right),\\
c_1&=\frac12 \cos\left( \frac{\pi}{2K'}{\rm Im}(u-u_1)\right),
\end{align}
with keeping the functional forms of Eqs.\,\eqref{eq:compositestate} and \eqref{eq:x2dependence}. Indeed, one can check that this configuration gives the correct topological charge ${\bf m}=(1,0,-1)$ and satisfies the periodic boundary conditions without the flavor symmetry twist. 
%the contribution of these cosine functions exactly cancel out the sign reversal in Eqs.\eqref{eq:phiperiodicity1} and \eqref{eq:phiperiodicity2} and the resulting given initial configuration satisfies the untwisted version of the periodic boundary conditions \eqref{eq:untwistedPBC2}.

\begin{figure}[h]
\centering
\includegraphics[width=10cm]{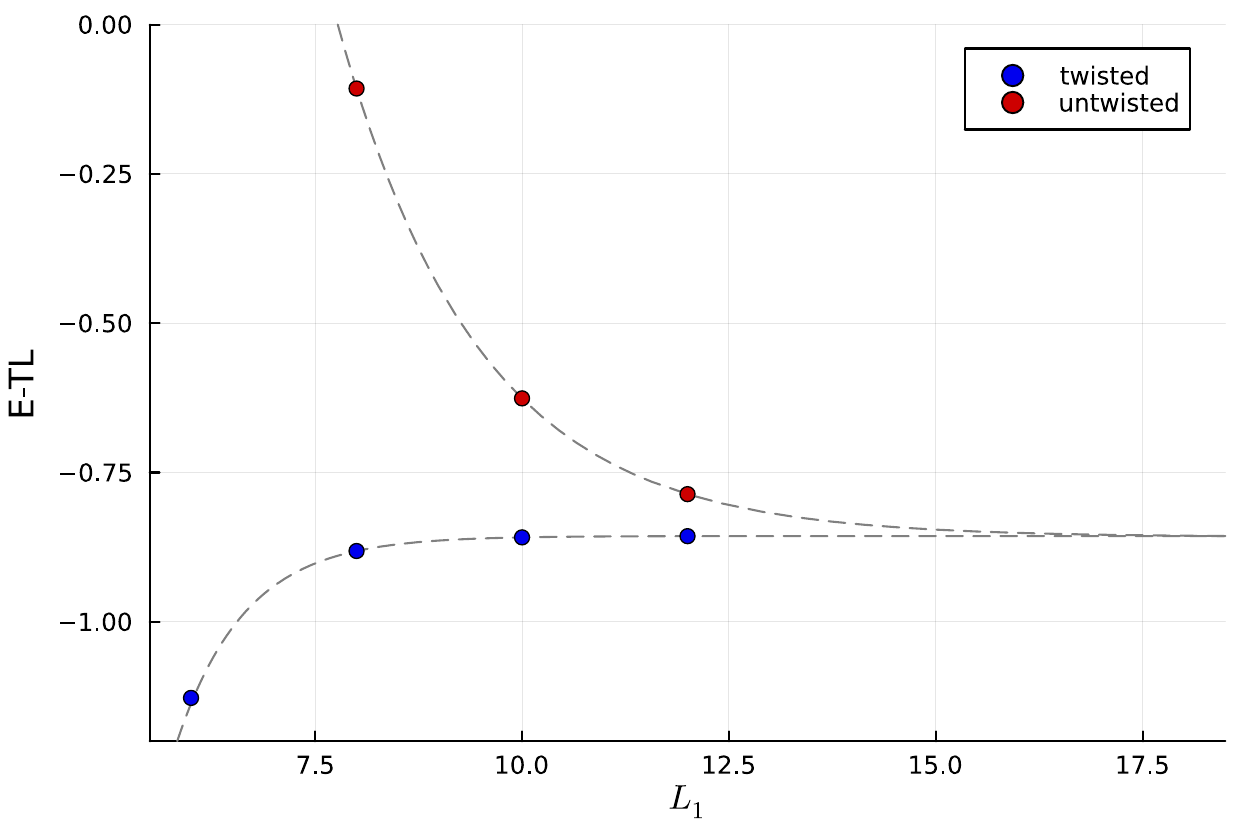}
\caption{$L_i$-dependence of subleading energy of the string junction for each fundamental domain of $T^3$, with the 
lattice spacing $a=1/4$ and keeping the ratio of the periods, $L_1:L_2:L_3=8:7:5$. Here the red dots represent the results for the untwisted boundary condition and the blue dots are for the twisted one discussed in the main text. The dashed curves interpolate the data points using the fitting functions of the form $f(L)=2M+A e^{-B L}$.}\label{fig:ELdependenceVS}
\end{figure}

Starting from these initial configurations,
we performed numerical calculations under the untwisted boundary conditions \eqref{eq:untwistedPBC21}-\eqref{eq:untwistedPBC23}, following the similar method described in  Sec.\,\ref{sec:NumericalResults}. The resulting configurations are qualitatively similar to those obtained under the twisted boundary conditions shown in Fig.\,\ref{fig:dencities}. 
The differences become apparent when we plot the total energy minus the contribution of the string tension $TL~(L=\sqrt{3}/2 L_1 + L_2)$, as shown in Fig.\,\ref{fig:ELdependenceVS}. 
The data for both the twisted and untwisted cases can be well approximated using fitting functions of the form $f(L)=2M+A e^{-B L}$.\footnote{
This factor $A e^{-B L}$ is introduced by roughly estimating the effects of the various interactions in this configuration.
The main contribution comes from the interaction of each string with its copies in the adjacent fundamental regions. 
The interactions are repulsive if they have the same $U(1)$ phases, and attractive if they are opposite, which is consistent with the result shown in Fig.\,\ref{fig:ELdependenceVS}.}
As expected, extrapolation of these functions to $L \rightarrow \infty$ yields approximately the same value:
 \begin{align}
M=-0.4286(4)  \quad {\rm for~} a=\frac14.
\end{align}
This result indicates that for large $L_i$, the configuration around the junction point is (almost) independent of the boundary conditions.

%Let us discuss the cases with other types of boundary conditions.
%First, consider the case where the Dirichlet boundary condition is imposed instead of the periodic boundary condition. There are infinite variations in the choice of coordination in the two-dimensional plane of the boundary, but if we choose a lump solution, i.e., a configuration in the section of an infinitely extending string solution, we do not need to consider the contribution from the boundary in the limit of infinity, and can expect the resulting junction mass to be the same value as in the above case.

%Care should be taken when considering Neumann boundary conditions.
%If the string penetrates the plane of the boundary, as in this case, the plane must be orthogonal to the string. Otherwise, the end points of the string on the boundary will continue to slide due to tension of the string and will not form a stable solution. If the shape of the boundary is chosen correctly, the solution should be exactly theoretically equivalent to the solution on the torus, since our model has CP symmetry and thus the mirror method can be applied.
%In this way, it is possible to form a solution containing only one junction point, but it is necessary to consider a suitable triangular-shaped region as the base space.

%%%%%%%%%%%%%%%%%%%%%%%%%%%
%\bibliographystyle{apsrev4-1}
\bibliographystyle{jhep}
\bibliography{references}

\end{document}